\documentclass{article}
\usepackage{times,fullpage,amsfonts}
\begin{document}

\newtheorem{thm}{Theorem}
\newtheorem{prop}[thm]{Proposition}
\newtheorem{cor}[thm]{Corollary}
\newtheorem{lemma}[thm]{Lemma}
\newtheorem{claim}[thm]{Claim}
\newtheorem{defn}{Definition}

\newcommand{\ket}[1]{|#1\rangle}
\newcommand{\Pl}{{\cal P}}
\newcommand{\A}{{\cal A}}
\newcommand{\proof}{\noindent {\bf Proof:}}
\newcommand{\qed}{\hfill QED \medskip}

\title{Proof of
security of quantum key distribution with two-way
classical communications}
\author{Daniel Gottesman\\
EECS: Computer Science Division\\
University of California\\
Berkeley, CA 94720\\
E-mail: gottesma@eecs.berkeley.edu
\and
Hoi-Kwong Lo\\
MagiQ Technologies Inc.\\
275 Seventh Avenue, 26th Floor\\
New York, NY 10001\\
E-mail: hoi\_kwong@magiqtech.com
}

\date{\today}
\maketitle
\begin{abstract}
Shor and Preskill have provided a simple proof of
security of the
standard quantum key distribution scheme by Bennett and Brassard
(BB84) by demonstrating a connection between key distribution and
entanglement purification protocols with one-way communications.  Here
we provide proofs of security of standard quantum key distribution
schemes, BB84 and the six-state scheme, against the most general
attack, by using the techniques of {\it two}-way entanglement
purification.  We demonstrate clearly the advantage of classical
post-processing with two-way classical communications over classical
post-processing with only one-way classical communications in
QKD. This is done by the explicit construction of a new protocol for
(the error correction/detection and privacy amplification of) BB84
that can tolerate a bit error rate of up to $18.9\%$, which is higher
than what {\it any} BB84 scheme with only one-way classical
communications can possibly tolerate.  Moreover, we demonstrate the
advantage of the six-state scheme over BB84 by showing that the
six-state scheme can strictly tolerate a higher bit error rate than
BB84. In particular, our six-state protocol can tolerate a bit error
rate of $26.4\%$, which is higher than the upper bound of $25\%$ bit
error rate for any secure BB84 protocol.  Consequently, our protocols
may allow higher key generation rate and remain secure over longer
distances than previous protocols.  Our investigation suggests that
two-way entanglement purification is a useful tool in the study of
advantage distillation, error correction, and privacy amplification
protocols.

\end{abstract}

\section{Introduction}
\label{Intro}
Quantum key distribution (QKD) \cite{bb84,ekert}
\footnote{The first paper on quantum cryptography was written by
Stephen Wiesner around 1970, but it remained unpublished until 1983
\cite{wiesner}. For a survey on quantum cryptography, see, for
example, \cite{pt}. For a review, see, for example, \cite{gisin}.}
allows two parties to communicate in absolute privacy in the presence
of an eavesdropper. Unlike conventional schemes of key distribution
that rely on unproven computational assumptions, the security of QKD
is guaranteed by the Heisenberg uncertainty principle of quantum
mechanics.  Much of the interest in QKD arises from the possibility of
near-term real-life applications, whereas most other potential uses of
quantum information remain remote.  QKD has been performed
experimentally over about 67~km of telecom fibers, and point-to-point
through open air at a distance of about 23.4~km.  There are also
proposals for key exchange from ground to satellites.  (See
\cite{pt,gisin} for discussions.)

Today's technologies fall short of full control and manipulation of
quantum states, so practical QKD protocols must use a much more
restricted set of operations. The best-known QKD protocol was
published by Bennett and Brassard in 1984 \cite{bb84}.  BB84 is a
simple ``prepare-and-measure'' protocol that can be implemented
without a quantum computer (see~\cite{nc,lps} for background on quantum
computation).  In a ``prepare-and-measure'' protocol, Alice simply
prepares a sequence of single-photon signals and transmits them to
Bob.  Bob immediately measures those signals; thus, no quantum
computation or long-term storage of quantum information is necessary,
only the transmission of single-photon states, which can be performed
through regular optical fibers.  Therefore, ``prepare-and-measure''
schemes are good candidates for near-term implementations of quantum
cryptography.

Of course, a theoretical description of a protocol is a mathematical
idealization.  Any real-life quantum cryptographic system is a complex
physical system with many degrees of freedom, and is at best an
approximation to the ideal protocol.  Proving the security of any
particular set-up is a difficult task, requiring a detailed model of
the apparatus.  Even a seemingly minor and subtle omission can be
fatal to the security of a cryptographic system.

Nevertheless, a number of important basic issues have been identified.
See, for example, \cite{norbert} for a discussion.  For instance, the
ideal theoretical version of BB84 uses a perfect single-photon source.
It is important to know whether an eavesdropper can in principle
exploit imperfect photon sources or other minor deviations from the
ideal model (such as channel loss or detector dark counts).
In this paper, we will not consider the issue of imperfections in the
source or detectors.  Instead, we will concentrate on the allowable
bit error rate in the channel, and show that it can be at least 26.4\%
for a ``prepare-and-measure'' scheme.

To prove the security of a protocol, one must specify clearly what
eavesdropping strategies are permissible.  In classical cryptography,
eavesdroppers are frequently given only a bounded amount of
computation.  Unfortunately, we do not, as yet, have a good grasp of
what can be done with a short quantum computation, and provable bounds
are elusive, even for classical computation.  Other assumptions are
similarly unreliable, so we resort to one of the most conservative
assumptions, unconditional security --- that is, security against the
most general attacks allowed by quantum mechanics.

As it turned out, proving unconditional security even for an idealized
system was very difficult.  More than a decade passed between the
original proposal for BB84 and the first general but rather complex
proof of security by Mayers~\cite{mayersqkd}, which was followed by a
number of other proofs~\cite{others,others2}.  Another approach to
proving the security is to start by considering protocols which are
less experimentally accessible.  In particular, Lo and
Chau~\cite{qkd}, building on the quantum privacy amplification idea of
Deutsch {\it et al.}~\cite{deutsch}, have proposed a conceptually
simpler proof of security.  However, the protocol proved secure has
the unfortunate drawback of requiring a quantum computer. Recently,
Shor and Preskill \cite{shorpre} have unified the techniques in
\cite{qkd} and \cite{mayersqkd} and provided a simple proof of
security of standard BB84.  (See also~\cite{squeezed} for a detailed
exposition of this proof.)\footnote{Mayers' and Shor-Preskill's proofs
make different assumptions. While Mayers' proof assumes that Alice's
preparation of the BB84 states is perfect, Shor and Preskill limit the
types of imperfections allowed in Bob's measurement apparatus. A proof
that takes into account more general imperfections remains to be
published.}

The idea of an entanglement purification protocol (EPP) \cite{BDSW}
plays a key role in Shor and Preskill's proof.  An EPP is a procedure
allowing Alice and Bob to create a small number of reliable EPR pairs
from a larger number of noisy pairs.  More specifically, Shor and
Preskill consider schemes for entanglement purification with a
classical side channel from Alice to Bob (one-way EPPs), which, by the
earlier work of Bennett, DiVincenzo, Smolin and Wootters
(BDSW)\cite{BDSW}, are mathematically equivalent to quantum
error-correcting codes (QECCs).

As noted by BDSW, EPPs involving two-way communications between two
parties can tolerate a substantially higher error rate than one-way
EPPs.  Those two-way EPPs are useful for the transmission of quantum
signals, but not their storage in a noisy memory, since in a two-way
EPP, the receiver Bob must send information to the sender Alice.

In this paper, we demonstrate that it is possible to create
``prepare-and-measure'' QKD schemes based on two-way EPPs, and that
the advantages of two-way EPPs can survive. More specifically, we
describe versions of BB84 and the six-state scheme \cite{bruss} (another
``prepare-and-measure'' scheme) using two-way communcations and prove
their security with allowed error rates substantially
higher than any previous
proofs.

Our results are significant for QKD for several reasons.  First, our
scheme can tolerate substantially higher bit error rates than all previous
protocols.  This may allow us to extend the distance of secure QKD and
increase the key generation rate.  Second, we demonstrate clearly the
advantage of using {\em two}-way classical communications in the
classical post-processing of signals in QKD.  In particular, for both
BB84 and the six-state scheme, our protocol tolerates a higher bit
error rate than any one-way post-processing method.  Third, our
results show rigorously that the six-state protocol can tolerate a
higher bit error rate than BB84.  These facts can help direct
experimentalists towards the most robust schemes for quantum key
distribution.

There are good conceptual reasons as well for studying two-way QKD.
The Shor and Preskill proof of security turns on the relationship
between classical error correction and privacy amplification and
QECCs.  EPPs have a close relationship to QECCs, but the detailed
connection between EPPs using one-way and two-way classical side
channels is not well understood \cite{BDSW}; in fact, very little is
known about two-way EPPs.  A study of two-way QKD elucidates the
relationship between the various aspects of quantum cryptography and
two-way EPPs. It may help to spur progress in both the theoretical
study of two-way EPPs and also their practical applications in a real
experiment. This is so because ``prepare-and-measure'' QKD schemes,
which we consider, can essentially be implemented in a real experiment
\cite{norbert}.  Furthermore, the study of two-way QKD can clarify
other proofs of security of QKD such as that due to Inamori
\cite{inamori,inamori6}, and may make the connection to earlier
studies of classical advantage distillation
\cite{maurer0,maurer1,maurer2}.%
\footnote{An important result in classical cryptography based on noisy
channels is that a two-way side channel can actually increase the
secrecy capacity of a noisy channel.  i.e., the secrecy capacity with
a two-way side channel, $C_2^s $, can be strictly greater than the
secrecy capacity with only a one-way side channel,
$C_1^{s}$. See~\cite{maurer0,maurer1,maurer2} for details. This is in sharp
contrast with Shannon's channel coding theorem which states that
two-way side channels do not increase channel capacity.  The process
of using two-way communications to share a secret in a way that is
impossible with only one-way communications is called ``advantage
distillation''.}

In section~\ref{s:protocol}, we present the BB84 and six-state
protocols and review known bounds on the bit error rates they
tolerate.  Section~\ref{s:entangle} reviews the necessary concepts
from the theory of quantum error-correcting codes and entanglement
purification protocols.  Even readers already familiar with these
subjects may wish to read sections~\ref{s:EPPQKD} and \ref{s:entangle}
to acquaint themselves with our terminology.
Section~\ref{s:shorpreskill} presents the Shor and Preskill proof of
security.  In section~\ref{s:naive}, we attempt a naive generalization
of the proof to two-way protocols, which fails in an instructive way.
In section~\ref{secmain} we present the main theorem: EPPs satisfying
the correct set of conditions can be made into secure
``prepare-and-measure'' QKD schemes with two-way communications.  An
example EPP satifying the conditions is presented in
section~\ref{EPPB}; variations of this EPP produce the achievable
error rates cited in this paper.  We prove the main theorem in
section~\ref{s:proof}.

\section{QKD protocols and bounds on performance}
\label{s:protocol}

\subsection{BB84 and the six-state scheme}

In the BB84 protocol for QKD, Alice sends a qubit (i.e., a quantum bit or
a two-level quantum system) in
one of four states to Bob.
The states $\ket{0}$ and $\ket{+} = (\ket{0} + \ket{1})/\sqrt{2}$
represent the classical bit $0$, while the states $\ket{1}$ and
$\ket{-} = (\ket{0} - \ket{1})/\sqrt{2}$ represent the bit $1$.  Alice
chooses one of these four states uniformly at random, and sends it to
Bob, who chooses randomly to measure in either the $\ket{0}$,
$\ket{1}$ basis (the ``$Z$'' basis) or the $\ket{+}$, $\ket{-}$ basis
(the ``$X$'' basis).  Then Alice and Bob announce the basis each of them
used for each state (but not the actual state sent or measured in that
basis), and discard any bits for which they used different bases.  The
remaining bits form the raw key, which will be processed some more
to produce the final key.

The six-state protocol is quite similar, but Alice sends one of six
states instead of one of four.  The four states from BB84 are used
(with the same meanings), plus the two states $(\ket{0} + i
\ket{1})/\sqrt{2}$ and $(\ket{0} - i \ket{1})/\sqrt{2}$, which
represent $0$ and $1$ in the ``$Y$'' basis.  Bob chooses to measure
randomly in the $X$, $Y$, or $Z$ basis, and again Alice and Bob
discard any bits for which they used different bases.  Thus, for the
six-state scheme, the raw key consists of one-third of the qubits
received on average, as opposed to one-half for
BB84;%
\footnote{Prepare-and-measure QKD schemes can be made more efficient
by employing a refined data analysis in which the bit error rates of
the sampled data of the various bases are computed separately and each
demanded to be small.  See \cite{eff,patent} for discussions and a
proof of the unconditional security of those efficient
prepare-and-measure QKD schemes.}
however, as we shall see, the six-state scheme remains secure under
noisier conditions.

Once they have produced the raw key, Alice and Bob select a sample of
sufficient size (assume one-half the total raw key for simplicity),
and publicly announce the values of those bits.  They compare and
calculate the fraction of bits which disagree; this is known as the
``bit error rate.''  The bit error rate gives an estimate of the
actual error rate for the remaining key bits.  If the bit error rate
is too high, Alice and Bob assume there is an eavesdropper and abort
the protocol.  Otherwise, Alice and Bob take their remaining bits and
may correct them using a classical error-correcting code: that is, Alice
announces her values for the parity checks of a classical linear code,
and Bob compares his values for the same parity checks to deduce the
locations of errors in the remaining key bits.  He corrects those
errors. Finally, Alice and Bob perform privacy amplification whose
goal is to remove the eavesdropper's information on the final key:
they choose some set of parities, and the final key bits are the
values of those parities.  After this procedure, provided the bit
error rate is not too high, the final key is supposed to be secure
against an eavesdropper Eve.

There are a few points about the protocols which deserve additional
comment.  First, all of Alice and Bob's classical communications occur
over a public channel, so Eve also has available to her any
information that was announced.  However, the classical channel should
be authenticated, so that Eve can only listen to it and not change it.
Second, after producing the raw key and before performing the error
test, Alice and Bob should agree on a random permutation to apply to
their raw key bits.  This simplifies the analysis, since Eve's attack
under these circumstances might as well be symmetric over all qubits
sent, and improves the tolerable bit error rate.  Third, the meaning
of ``security'' for this protocol is slightly subtle: for any attack
chosen by Eve, either she will be detected, except with probability
exponentially small in some security parameter, $r$, or, with
probability exponentially close to 1, she will have an exponentially
small amount of information, in some security parameter, $s$, about
the final key.
A QKD scheme is efficient if the resources (in terms of the number of
qubits sent, amount of computational power, etc) required for its
implementation are at most polynomial in the security parameters. 
For simplicity, it is quite common to take the security parameters to
be $n$, the total number of qubits sent. As discussed
in \cite{eff}, other choices of the security parameters are perfectly
acceptable.

\subsection{Known bounds on the performance of QKD}
 
There are a number of upper and lower bounds known for the allowable
bit error rate for these two protocols.  In table~\ref{table:bounds}, we
summarize the bounds for BB84 and the six-state scheme.
\begin{table}
\centering
\begin{tabular}{lcc}
\multicolumn{3}{c}{BB84} \\
\\
                      &  one-way &  two-way  \\
Upper bound           &  $14.6\%$  &  $1/4$   \\
Lower bound           &  $11.0\%$  &  $18.9\%$ \\
\\
\multicolumn{3}{c}{Six-state Scheme} \\
\\
                      &  one-way  &  two-way  \\
Upper bound           &  $1/6$    &  $1/3$    \\
Lower bound           &  $12.7\%$ &  $26.4\%$ 
\end{tabular}
\caption{Bounds on the bit error rate for BB84 and the six-state
scheme using one-way and two-way classical post-processing.  The lower
bounds for two-way post-processing, $18.9\%$ for BB84 and
$26.4 \%$ for the six-state scheme, come from the current work.}
\label{table:bounds}
\end{table}
The tables give bounds for schemes that use one-way and two-way
classical communications during the post-processing phase.  The upper
bounds are derived by considering some simple individual attacks,
and determining when these attacks can defeat QKD.  The lower bounds
come from protocols that have been proved secure.  For both BB84 and
the six-state scheme, our new lower bounds for two-way classical
post-processing schemes are substantially better than the upper bounds
for schemes with one-way classical post-processing. Therefore,
our results demonstrate clearly that our schemes can tolerate
higher bit error rates than
any possible schemes with only one-way classical post-processing can.

The upper bounds for one-way post-processing come from attacks based
on optimal approximate cloning machines~\cite{fuchs,cirac,bechmann}.
Although perfect cloning of an unknown quantum state
is strictly forbidden by the uncertainty principle of
quantum mechanics, approximate cloning is possible.
Optimal approximate cloning has recently been experimentally
demonstrated \cite{lamas}.
More specifically,
Eve intercepts all of Alice's signals from the quantum channel.  Using
the appropriate optimal cloner, Eve can generate two equally good
approximate copies of the original signal.
In the case of BB84, the resulting bit error rate in a single copy is
about $14.6\%$ \cite{fuchs,cirac}, and it is $1/6$ for
the six-state scheme \cite{bechmann}. 
Eve then
keeps one copy herself and sends the second copy to Bob.  With only
one-way classical processing, Bob is not allowed to send classical
signals to Alice.\footnote{If one allows Bob to send classical messages
to Alice only (but not from Alice to Bob),
in the context of coherent state QKD,
it is known that such backward one-way communications
can actually help to beat the approximate cloning attack.
However, the issue of unconditional security remains open.
See \cite{grangier}
for details.}
Therefore, Bob and Eve are in a completely
symmetric situation: if Bob can generate a key based on subsequent
classical transmissions from Alice, Eve must be able to do the
same.  Therefore, at this error rate ($14.6\%$ or $1/6$), the QKD scheme
must be insecure with one-way post-processing.

The upper bounds for two-way post-processing come from an intercept
and resend eavesdropping strategy.  Eve intercepts each qubit sent by
Alice.  She chooses to measure in a random basis from the appropriate
list ($X$, $Z$ for BB84 or $X$, $Y$, $Z$ for the six-state scheme).
She records her measurement outcome and prepares a single photon in
the polarization given by her measurement outcome and re-sends such a
photon to Bob. Note that whatever Bob can do from this point on can
be simulated by a classical random variable {\em prepared} by Eve, who
has a classical record of it, and a local random number generator
possessed by Bob. Therefore, secure QKD is impossible even with
two-way classical communications between Alice and Bob.  For BB84, the
intercept and resend strategy gives an error rate of $25\%$: half the
time Eve has chosen the correct basis, so there is no error, and half
the time she has chosen the wrong basis, in which case there is a
$50\%$ chance of an error, for a net error rate of $1/4$.  For the
six-state scheme, intercept and resend gives an error rate of $1/3$:
Eve has the correct basis only $1/3$ of the time, and the remaining
$2/3$ of the time, she has a $50\%$ chance of introducing an error.

The lower bounds in table~\ref{table:bounds} come from proofs of
security.  The Shor and Preskill proof shows that QKD with one-way
communications can be secure with data rate at least $1 - 2h(p)$,
where $p$ is the bit error rate and $h(x) = -x \log_2 x - (1-x) \log_2
(1-x)$ is the Shannon entropy.  This reaches $0$ when $p$ is about
$11.0\%$.  For the six-state scheme, this result has been slightly
improved~by one of us (H-K. Lo) \cite{six} to allow secure
QKD up to a bit error rate of
about $12.7\%$.%
\footnote{The result in \cite{six} makes use of the non-trivial mutual
information between the bit-flip and phase error syndromes, and of the
degenerate codes studied by DiVincenzo, Shor and Smolin \cite{dss}.}
With two-way communications during post-processing, Shor and
Preskill's result and Lo's result remain the best prior results. (Lo's
result is marginally better than Inamori's result~\cite{inamori6} for
the six-state scheme, which requires two-way classical
post-processing.) In this paper, we present significant improvements
on both those lower bounds.

\subsection{EPP schemes for QKD}
\label{s:EPPQKD}

For our proof of security, it will be helpful to consider another
class of scheme based on EPPs (which are described in more detail in
section~\ref{s:entangle}).  For these QKD schemes, which we will refer
to as {\em EPP schemes} or {\em EPP protocols},\footnote{``EPP
protocol'' sounds redundant since the second ``P'' in ``EPP'' also
stands for ``protocol.''  However, it is not really redundant, since
the full phrase is short for ``quantum key distribution protocol based
on an entanglement purification protocol.''}  Alice prepares a number
of EPR pairs $\ket{\Psi^+} = (\ket{00} + \ket{11})/\sqrt{2}$.  On the
second qubit of each pair, Alice then performs a random rotation
chosen either from the set $I$, $H$ or the set $I$, $T$, $T^2$.  $I$
is the identity operation, $H$ is the Hadamard transform, which swaps
states in the $X$ and $Z$ bases, and $T$ is a unitary operation which
takes states in the $X$ basis to the $Y$ basis, states in the $Y$
basis to the $Z$ basis, and states in the $Z$ basis to the $X$ basis.

We will refer to the first case (with $I$ and $H$) as the {\em
two-basis} EPP protocol, and the second case (with $I$, $T$, and
$T^2$) as the {\em three-basis} EPP protocol.  The two-basis scheme
will produce a protocol related to BB84, while the three-basis scheme
produces a protocol related to the six-state scheme.  We can also
consider {\em efficient} schemes in which the rotations are not
performed with equal probabilities.  These produce efficient BB84 and
six-state schemes~\cite{eff,patent}, which have a higher rate of key
generation per qubit transmitted.

After performing the rotation, Alice sends the second qubit of each
pair to Bob.  When Bob acknowledges receiving the transmission, Alice
announces which rotation she performed for each pair.  Bob reverses
this rotation.  Then Alice and Bob agree on a random permutation of
the EPR pairs, and select a subset (half of the pairs by default) to
measure (in the $Z$ basis) to test for errors.  They compare the
results of the test, and abort if the error rate is too high.  If not,
Alice and Bob perform an entanglement purification protocol to extract
good entangled pairs.  Then they measure (again in the $Z$ basis) the
remaining pairs and use the result as their secret key.

The security proofs we review in section~\ref{s:shorpreskill} show
that the security of BB84 and the six-state scheme can be reduced to
the security of the above EPP schemes using appropriate entanglement
purification protocols.  The protocols that lead to traditional
prepare-and-measure
one-way post-processing schemes are EPPs using just one-way
communications; in this paper, we present two-way post-processing
schemes that arise from EPPs with two-way classical communications.

\section{Entanglement purification and quantum error correction}
\label{s:entangle}

Suppose Alice and Bob are connected by a noisy quantum channel (and
perhaps also a noiseless classical channel). Entanglement purification
provides a way of using the noisy quantum channel to simulate a
noiseless one. More concretely, suppose Alice creates $N$ EPR pairs
and sends half of each pair to Bob. If the quantum communication
channel between Alice and Bob is noisy (but stationary and
memoryless), then Alice and Bob will share $N$ imperfect EPR pairs,
each in the state $\rho$. They may attempt to apply local operations
(including preparation of ancillary qubits, local unitary
transformations, and measurements) and classical communications (LOCCs)
to purify the $N$ imperfect EPR pairs into a smaller number, $n$, EPR
pairs of high fidelity.  This process is called an entanglement
purification protocol (EPP) and was first studied by Bennett,
DiVincenzo, Smolin and Wootters (BDSW)~\cite{BDSW}.

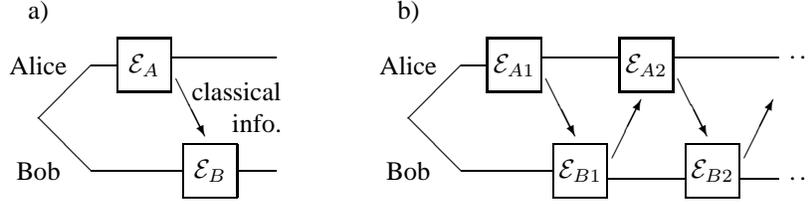
\begin{figure}

\centering

\begin{picture}(305,90)

\put(5,72){\makebox(10,16){a)}}

\put(10,40){\line(1,1){20}}
\put(10,40){\line(1,-1){20}}
\put(0,52){\makebox(20,16){Alice}}
\put(0,12){\makebox(20,16){Bob}}

\put(30,60){\line(1,0){10}}
\put(40,50){\framebox(20,20){${\cal E}_A$}}
\put(30,20){\line(1,0){35}}

\put(62,55){\vector(1,-2){11}}
\put(65,10){\framebox(20,20){${\cal E}_B$}}

\put(60,63){\line(1,0){40}}
\put(85,20){\line(1,0){15}}

\put(72,44){\makebox(30,12)[r]{classical}}
\put(72,32){\makebox(30,12)[r]{info.}}

\put(145,72){\makebox(10,16){b)}}

\put(150,40){\line(1,1){20}}
\put(150,40){\line(1,-1){20}}
\put(140,52){\makebox(20,16){Alice}}
\put(140,12){\makebox(20,16){Bob}}

\put(170,60){\line(1,0){10}}
\put(180,50){\framebox(20,20){${\cal E}_{A1}$}}
\put(170,20){\line(1,0){35}}

\put(202,55){\vector(1,-2){11}}
\put(205,10){\framebox(20,20){${\cal E}_{B1}$}}

\put(200,63){\line(1,0){30}}
\put(227,25){\vector(1,2){11}}
\put(230,50){\framebox(20,20){${\cal E}_{A2}$}}

\put(225,17){\line(1,0){30}}
\put(252,55){\vector(1,-2){11}}
\put(255,10){\framebox(20,20){${\cal E}_{B2}$}}

\put(250,63){\line(1,0){40}}
\put(275,17){\line(1,0){15}}
\put(295,59){\makebox(10,8){\ldots}}
\put(295,13){\makebox(10,8){\ldots}}

\put(277,25){\vector(1,2){11}}

\end{picture}

\caption{a) A 1-EPP.  Alice performs some unitary operations and
measurements, then makes a transmission to Bob, who performs another
unitary transformation, possibly based on Alice's classical
transmission.  b) A 2-EPP.  Alice and Bob alternate local operations
and classical transmissions.  Each operation can depend on the
contents of earlier transmissions.  The procedure can extend
indefinitely.}
\label{fig:EPP}
\end{figure}

One way to classify EPPs is in terms of what type of classical
communications they require.  Figure~\ref{fig:EPP}a shows the
structure of EPPs that can be implemented with only one-way classical
communications from Alice to Bob, known as one-way EPPs or 1-EPPs.
Figure~\ref{fig:EPP}b shows the structure of EPPs requiring two-way
classical communications, known as two-way EPPs or 2-EPPs.

Typically, a 1-EPP will consist of Alice measuring a series of
commuting operators and sending the measurement result to Bob.  Bob
will then measure the same operators on his qubits.  If there is no
noise in the channel, Bob will get the same results as Alice, but of
course when noise is present, some of the results will differ.  From
the algebraic structure of the list of operators measured, Bob can
deduce the location and nature of the errors and correct them.
Unfortunately, the process of measuring EPR pairs will have destroyed
some of them, so the resulting state consists of fewer EPR pairs than
Alice sent.

As noted by BDSW, a 1-EPP is mathematically equivalent to a quantum
error-correcting code (see~\cite{qecc,nc} for background on QECCs).
Instead of measuring a series of operators and transmitting the
results, Alice instead encodes Bob's qubits into a particular
predetermined eigenspace of the list of operators.  Then when Bob
receives the qubits, he can measure the same list of operators,
telling him the error syndrome for the QECC given by that subspace.
For instance, if the channel only produces bit flip errors, Alice can
encode Bob's state using a random coset of a classical linear code,
and then Bob measures the parity checks for that code.  He determines
what error the channel introduced by calculating how the coset has
changed since Alice's transmission.

Two-way EPPs can be potentially more complex, but frequently have a
similar structure.  Again, Alice and Bob measure a set of identical
operators.  Then they compare their results, discard some EPR pairs,
and together select a new set of operators to measure.  An essential
feature of a two-way EPP is that the subsequent choice of measurement
operators may depend on the outcomes of previous measurements.  This
process continues for a while until the remaining EPR pairs have a low
enough error rate for a 1-EPP to succeed. Then, a 1-EPP is applied.

Unfortunately, not all EPPs are suitable for making a
prepare-and-measure QKD protocol.  The next few definitions are
designed to set the stage for the detailed sufficient conditions in
our main theorem.  We will, for instance, primarily be interested in a
restricted class of EPPs which involve the measurement of Pauli
operators. The best studied EPPs can all be described in the
``stabilizer'' formulation, which employs Pauli operators
extensively.  Other EPPs might still be useful for QKD, but are
less well studied.

\begin{defn}
A {\em Pauli operator} acting on $n$ qubits is a tensor product of
individual qubit operators that are of the form $I$ (the identity), $X
= \pmatrix{0 & 1 \cr 1 & 0}$, $Y = \pmatrix{0 & -i \cr i & \ 0}$, and
$Z = \pmatrix{1 & \ 0 \cr 0 & -1}$.  An {\em $X$-type} operator is a
tensor product of just $I$s and $X$s, and a {\em $Z$-type} operator is
a tensor product of just $I$s and $Z$s.
\end{defn}

Note that the states in section~\ref{s:protocol} described as being in
the $X$, $Y$, or $Z$ bases are in fact eigenstates of the operators
$X$, $Y$, and $Z$.  A CSS code involves measuring just $X$-type and
$Z$-type Pauli operators.  Also, note that any pair of $X$, $Y$, and
$Z$ anticommute with each other (so, for instance, $XZ = -ZX$).
Finally, note that all Pauli operators have only eigenvalues $+1$ and
$-1$.  Classical linear error-correcting codes can be understood as a
measurement of a series of just $Z$-type operators: the eigenvalue of
a $Z$-type operator is the parity of bits on which the operator acts
as $Z$.  (For instance, measuring $Z \otimes I \otimes Z$ gives the
parity of the first and third bits.)

When dealing extensively with Pauli operations, it is helpful to also
look at a more general class of operators which interact well with
Pauli operations.  

\begin{defn}
A unitary operation belongs to the {\em Clifford group} if it
conjugates Pauli operators into other Pauli operators.
\end{defn}

Thus, a Clifford group operation will map eigenstates of a Pauli
operation into eigenstates of another Pauli operation.  For instance,
CNOT and $H$ are both Clifford group operations.  (In fact, the Clifford
group is generated by CNOT, $H$, and the phase gate $\ket{0} \rightarrow
\ket{0}$, $\ket{1} \rightarrow i \ket{1}$.)

\begin{defn}
We say an EPP (one-way or two-way) is {\em symmetric} if it can be
described with a set of operators $\{M_\mu\}$, plus unitary decoding
operations $U_{\mu} \otimes (P_{\mu} U_{\mu})$.  Each operator $M_\mu$
describes a measurement that may be made at some point in the
protocol; the index $\mu$ describes a history of outcomes of earlier
measurements as a string of $0$s and $1$s.  On the history $\mu$,
Alice performs the measurement $M_\mu$ on her side, and Bob performs
the measurement $M_\mu$ on his side.  (They always perform the same
sequence of measurements, thus the name ``symmetric.'')  They then
update the history $\mu$ by appending the parity of their two
measurement outcomes ($0$ for the same outcome, $1$ for opposite
outcomes).  The protocol begins with each person measuring the
operator $M_{\emptyset}$.  Each time the history is updated, Alice and
Bob measure the operator corresponding to the new value of $\mu$, and
again update the history according to the result.  When there is no
$M_\mu$ for the current history, Alice performs the operation
$U_{\mu}$ and Bob performs the operation $P_{\mu} U_{\mu}$, and the
protocol terminates.
\end{defn}

\begin{figure}

\centering

\begin{picture}(220,90)

\put(30,40){\line(1,1){20}}
\put(30,40){\line(1,-1){20}}
\put(20,52){\makebox(20,16){Alice}}
\put(20,12){\makebox(20,16){Bob}}

\put(50,60){\line(1,0){10}}
\put(60,50){\framebox(20,20){$M_{\emptyset}$}}
\put(50,20){\line(1,0){10}}
\put(60,10){\framebox(20,20){$M_{\emptyset}$}}

\put(85,50){\line(1,-1){10}}
\put(85,30){\line(1,1){10}}
\put(95,40){\line(1,0){10}}
\put(105,40){\vector(1,1){10}}
\put(105,40){\vector(1,-1){10}}
\put(96,40){\makebox(8,8){$r$}}

\put(80,60){\line(1,0){40}}
\put(120,50){\framebox(20,20){$M_{r}$}}
\put(80,20){\line(1,0){40}}
\put(120,10){\framebox(20,20){$M_{r}$}}

\put(145,50){\line(1,-1){10}}
\put(145,30){\line(1,1){10}}
\put(155,40){\line(1,0){10}}
\put(165,40){\vector(1,1){10}}
\put(165,40){\vector(1,-1){10}}
\put(156,40){\makebox(8,8){$s$}}

\put(140,60){\line(1,0){40}}
\put(180,50){\framebox(20,20){$M_{rs}$}}
\put(140,20){\line(1,0){40}}
\put(180,10){\framebox(20,20){$M_{rs}$}}

\put(200,60){\line(1,0){10}}
\put(200,20){\line(1,0){10}}
\put(210,56){\makebox(10,8){\ldots}}
\put(210,16){\makebox(10,8){\ldots}}

\end{picture}
\caption{Structure of a symmetric EPP.  Alice and Bob measure the same
sequence of operators.  $r$ and $s$ are the parities of the outcomes
of Alice's and Bob's measurements of $M_{\emptyset}$ and $M_r$,
respectively.}
\label{fig:symmetric}
\end{figure}
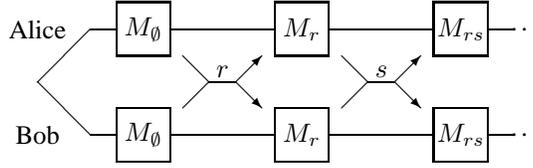

Figure~\ref{fig:symmetric} shows a symmetric EPP.  See also
section~\ref{s:tree} and figure~\ref{fig:tree} for another
representation.

Note that if the history $\nu$ is an extension of the history $\mu$
(i.e., it is $\mu$ with additional bits appended), the operators
$M_\mu$, $M_\nu$ should commute for the EPP to be realizeable using
local operations and no additional resources.  On the other hand, for
two different extensions, $\nu_1$ and $\nu_2$, of the same history
$\mu$, the corresponding operators $M_{\nu_1}$ and $M_{\nu_2}$ do {\em
not} need to commute. This is because Alice and Bob never need to
measure both operators for the same state.

For a 2-EPP, the commutation requirement is the only constraint on the
$M$s.  For a 1-EPP, we also require that the operators $M_\mu$ depend
only on the length of $\mu$ (i.e., how many measurements have been
made so far) and not the precise history.  This is because in a 1-EPP,
Alice cannot learn Bob's measurement outcomes and therefore cannot
know the exact value of the history $\mu$.

The final operation $U_{\mu} \otimes P_{\mu} U_{\mu}$ serves two
purposes.  First of all, the measurements have determined a good deal
of information about the state of the system, and we must disentangle
that from the residual Bell states.  Second, it acts to correct,
discard, or otherwise eliminate any errors identified by the
measurements.  For instance, if the EPP locates pairs with errors, but
does not identify what kind of errors are present, the final operation
$U_{\mu}$ would likely permute the qubits to move the errors to a
standard set of locations, which are then discarded.  It is convenient
to separate the decoding operation into two parts: $U_{\mu}$, which is
performed by both people and represents decoding and discarding bad
EPR pairs, and $P_{\mu}$, performed just by Bob, which represents
correcting EPR pairs which will be kept.  In practice, it is often
easier to specify an EPP by including unitary operations in between
measurements as well as at the end of the protocol, but this is an
equivalent definition, since the measurement operators $M_\mu$ can
instead be defined to take the change of basis into account.  Notice
that in a 1-EPP, $U_{\mu}$ cannot depend on $\mu$, whereas $P_{\mu}$
invariably will --- otherwise there would be no way to correct any
errors discovered in the course of the protocol.

\begin{defn}
A symmetric EPP is a {\em stabilizer} EPP if all measurements $M_\mu$
are of eigenspaces of Pauli operations, the decoding operation
$U_{\mu}$ is a Clifford group operation, and the correction operation
$P_{\mu}$ is a Pauli operation.  For a 1-EPP, we again make the
restriction that $U_{\mu} = U$ is independent of $\mu$.  A stabilizer
EPP is {\em CSS-like} if all $M_\mu$s are $X$-type or $Z$-type Pauli
operations, and $U_{\mu}$ involves only CNOTs.
\end{defn}

Stabilizer 1-EPPs can be thought of as another guise of stabilizer
quantum error-correcting codes.  The measurements $M_{\mu}$ correspond
to the generators of the code stabilizer.  $U_{\mu}$ is the decoding
operation, which for a stabilizer code is always from the Clifford
group, and $P_{\mu}$ corrects the Pauli errors that have occurred.
CSS-like 1-EPPs correspond to the class of CSS codes; since they are
based on classical linear codes, the decoding only needs CNOT
gates.

The same intuition applies to the case of 2-EPPs.  The condition that
decoding only needs CNOT means intuitively that the encoded $Z$
operation is, in fact, also of $Z$-type; that is, it can be written as
a tensor product of $Z$ operators.  The final correction operation
$P_{\mu}$ is a Pauli operator because the error syndrome (disclosed in
the two-way classical communication) should contain enough information
to identify which Pauli error has occurred in the quantum
channel.

The EPPs we will consider in this paper are all CSS-like EPPs.  In
fact, we will need to consider Alice and Bob choosing a random EPP out
of a family of similar EPPs, but this does not produce any further
intrinsic complications.  For simplicity, we may describe EPPs that
involve Clifford group or Pauli group operations in the middle of the
series of measurements instead of the end, but this does not affect
the definition at all; these EPPs can be rewritten to conform to the
above definition of stabilizer or CSS-like EPPs.

\subsection{A Tree Diagram Representation}
\label{s:tree}

The series of operators measured in a stabilizer 1-EPP or 2-EPP can be
represented using a tree diagram representation.\footnote{We thank
David DiVincenzo and Debbie Leung for suggesting the tree diagram
representation.}  Each vertex is labelled by an operator $M_\mu$ that
could be measured during the EPP. Each edge is labelled with one or
more possible outcomes of the previous measurements.  The edges are
directed from the root of the tree (labelled by $M_{\emptyset}$)
towards the leaves (labelled with $M_\mu$ for $\mu$ of maximal
length), representing the time-ordering of the measurements.

Given a tree diagram of the above form, we can read off the structure
of the EPP.  We start at the root of the tree, which is labelled by
measurement $M_{\emptyset}$.  We note the outcome and follow the edge
which is labelled by that outcome.  Then we perform the measurement
which labels the new vertex, and follow the edge corresponding to the
outcome of that measurement.  We repeat this process until we reach
the bottom of the tree, at which point we perform the appropriate
unitary operation $U_{\mu} \otimes P_{\mu} U_{\mu}$.  Each history
$\mu$ corresponds to a path through the tree.

\begin{figure}

\centering

\begin{picture}(380,170)

\put(5,152){\makebox(10,16){a)}}

\put(10,134){\makebox(60,12){$M_{\emptyset} = 
X \otimes Z \otimes Z \otimes X \otimes I$}}
\put(10,94){\makebox(60,12){$M_{r} = 
I \otimes X \otimes Z \otimes Z \otimes X$}}
\put(10,54){\makebox(60,12){$M_{rs} = 
X \otimes I \otimes X \otimes Z \otimes Z$}}
\put(10,14){\makebox(60,12){$M_{rst} = 
Z \otimes X \otimes I \otimes X \otimes Z$}}

\multiput(40,132)(0,-40){3}{\vector(0,-1){24}}

\put(175,152){\makebox(10,16){b)}}

\put(230,134){\makebox(60,12){$M_{\emptyset} =
X \otimes X \otimes X \otimes X$}}
\put(155,94){\makebox(60,12){$M_0 = Z \otimes Z \otimes Z \otimes Z$}}
\put(300,94){\makebox(60,12){$M_1 = X \otimes X \otimes I \otimes I$}}
\put(110,54){\makebox(60,12){stop}}
\put(195,54){\makebox(60,12){$M_{01} = Z \otimes I \otimes I \otimes Z$}}
\put(300,54){\makebox(60,12){$M_{1r} = X \otimes I \otimes X \otimes I$}}
\put(210,14){\makebox(60,12){stop}}
\put(300,14){\makebox(60,12){stop}}

\put(260,132){\vector(-3,-1){72}}
\put(205,117){\makebox(12,12){$0$}}
\put(260,132){\vector(3,-1){72}}
\put(307,117){\makebox(12,12){$1$}}

\put(190,92){\vector(-2,-1){48}}
\put(152,77){\makebox(12,12){$0$}}
\put(190,92){\vector(2,-1){48}}
\put(218,77){\makebox(12,12){$1$}}
\put(330,92){\vector(0,-1){24}}

\put(240,52){\vector(0,-1){24}}
\put(330,52){\vector(0,-1){24}}

\end{picture}

\caption{a) The tree diagram representation of a 1-EPP.  The sequence
of operators is fixed, so there is no branching.  The 1-EPP shown
corresponds to the 5-qubit QECC.  b) The tree diagram of a 2-EPP.
Future operators may depend on the outcome of a measurement, allowing
a branched tree.  When the tree branches, edges are labelled by the
outcome of the previous measurement.  When it does not branch, no
label is needed.  Note that the tree does not need to branch
uniformly, or even have uniform depth.  The EPP in part b) is
CSS-like; the EPP in part a) is not.}
\label{fig:tree}
\end{figure}
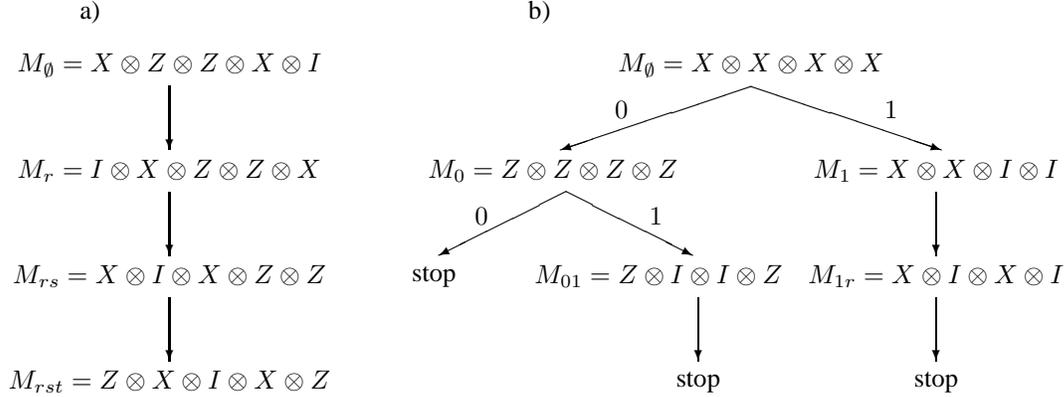

For any 1-EPP, the sequence of measurements does not depend on the
outcome of any measurement.  Therefore, a 1-EPP can be represented by
a straight (directed) line (figure~\ref{fig:tree}a).  On the other
hand, in a 2-EPP, the choice of measurement $M_\mu$ at any step $i$
can depend on the outcome of an earlier measurement $M_\nu$.  This
corresponds to a branch in the tree at step $i$ (see
figure~\ref{fig:tree}b).

\section{The Shor and Preskill Security Proof}
\label{s:shorpreskill}

Next, we give the Shor and Preskill proof \cite{shorpre} of security
of BB84.  See also \cite{squeezed} for a more detailed version.  Shor
and Preskill's proof begins by following Lo and Chau's proof
\cite{qkd} of the security of a scheme using EPPs, and then shows that
the security of BB84 follows from the security of the EPP scheme.

As noted before, in the EPP scheme, Alice creates $N$ EPR pairs and
sends half of each to Bob.  Alice and Bob then test the error rates in
the $X$ and $Z$ bases on a randomly chosen subset of $m$ pairs.  If
the error rate is too high, they abort; otherwise, they perform an EPP
$C$ on the remaining $N-m$ pairs. Finally, they measure (in the $Z$
basis) each of the $n$ EPR pairs left after $C$, producing a shared
random key about which, they hope, Eve has essentially no information.

\subsection{Noisy Quantum Channels and Eavesdropping Strategies}

All of the QKD protocols we consider will take place over a noisy
quantum channel, even when there is no eavesdropper present.  We shall
be primarily interested in a special class of quantum channels known
as Pauli channels.

\begin{defn}
A {\em quantum channel} is any superoperator which acts on transmitted
qubits.  A {\em Pauli channel} $(\Pl_i, q_i)$ applies the Pauli
operation $\Pl_i$ with probability $q_i$ (so we require $\sum q_i =
1$).  An {\em uncorrelated Pauli channel} $(q_X, q_Y, q_Z)$ applies a
random Pauli operator independently on each qubit sent through the
channel.  It applies $X$ with probability $q_X$, $Y$ with probability
$q_Y$, $Z$ with probability $q_Z$, and $I$ with probability
$1-q_X-q_Y-q_Z$.
\end{defn}

From the perspective of Alice and Bob, noise in the channel could have
been caused by an eavesdropper Eve.  We will need to consider two
types of eavesdropping strategy by Eve.  The first strategy, the joint
attack, is the most general attack allowed by quantum mechanics.

\begin{defn}
In a {\em joint attack} by Eve, Eve has a quantum computer.  She takes
all quantum signals sent by Alice and performs an arbitrary unitary
transformation involving those signals, adding any additional ancilla
qubits she cares to use.  She keeps any part of the system she desires
and transmits the remainder to Bob. She listens to the public
discussion (for error correction/detection and privacy amplification)
between Alice and Bob before finally deciding on the measurement
operator on her part of the system.

\end{defn}

The joint attack allows Eve to perform any quantum operation on the
qubits transmitted by Alice.  For the security proof, we shall also
consider a Pauli attack.

\begin{defn}
A {\em Pauli attack} by Eve is a joint attack where the final
operation performed on the transmitted qubits is a general Pauli
channel.
\end{defn}

\subsection{EPP protocols are secure}
\label{s:EPPproof}

In this subsection, we will show that the EPP protocols described in
section~\ref{s:protocol} are secure.  The argument is essentially that
of~\cite{qkd}.  First, what do we mean by ``secure?''

\begin{defn}
\label{defn:secure}
A QKD protocol to generate $n$ key bits is {\em correct} if, for any
strategy used by Eve, either Alice and Bob will abort with high
probability or, with high probability, Alice and Bob will agree on a
final key $k$ which is chosen nearly uniformly at random.  The
protocol is {\em secure} if, for any strategy used by Eve, either
Alice and Bob will abort with high probability or Eve's information
about the key will be at most $\exp(-s)$ for some security parameter,
$s$.  In all cases, ``with high probability'' means with probability
at least $1-\exp(-r)$ for some security parameter, $r$.  The resources
required for the implementation of a QKD scheme must be at most
polynomial in $r$ and $s$.  For simplicity, in what follows, we will
consider the case where $r=s$ and call it simply the security
parameter.
\end{defn}

Naively, one might consider a security requirement of the form
$I_{eve} < \delta n$, where $I_{eve}$ is the eavesdropper's mutual
information with the final key and $n$ is the length of the final key.
However, such a definition of security is too weak, since it allows
Eve to learn a few bits of a long message.  For instance, the
eavesdropper may know something about the structure of the message
that Alice is going to send to Bob.  Imagine that the last few
characters of the message contain the password for launching a nuclear
missile. In that case, Eve could compromise the security of the
message by concentrating her information on the last few bits.

Another naive definition of security would be to require that
$I_{eve} < e^{ - \alpha n}$ for any eavesdropping strategy.
Unfortunately, such a definition of security is too strong to be
achievable. For instance, Eve can simply replace the signal
prepared by Alice by sending Bob some signals with specific polarizations
prepared by herself. Such an eavesdropping attack is highly
unlikely to pass the verification test (by producing a small
error rate). However, in the unlikely event that it does
pass the verification test, Eve will have perfect information on
the key shared between Alice and Bob, thus violating the
security requirement $I_{eve} < e^{ - \alpha n}$.

In fact, even the definition we give is probably not strong enough for
some purposes: Eve can retain a {\em quantum} state at the end of the
protocol, and the security definition should refer to that rather than
bounding her {\em classical} information about the key.  For instance,
a better definition is: for any eavesdropping strategy, either Eve
will almost surely be caught, or, for any two final values of the key,
Eve's residual density matrices after the protocol concludes will have
high fidelity to each other.  That is, Eve cannot reliably distinguish
between any pair of values of the key.  We do not prove the stronger
definition in this paper.

The question of defining security for quantum cryptography in a way
that enables us to prove composibility of protocols remains an
important open problem.  For this paper, however, we simply use
definition~\ref{defn:secure}.

Our method will be to relate the security of BB84 and the six-state
scheme to the security of EPP schemes, and we wish to say that when
the EPP schemes are secure, so are the ``prepare-and-measure''
schemes.

\begin{defn}
Suppose QKD protocol $\beta$ is correct and secure, with a security
parameter $p$.  Then QKD protocol $\alpha$ is said to have security
{\em similar} to protocol $\beta$ when $\alpha$ is also correct and
secure, and its security parameter $q$ is polynomially related to $p$.
Furthermore, protocol $\alpha$ should abort at a given bit error rate
only if protocol $\beta$ also aborts at that bit error rate.
\end{defn}

To prove the security of EPP protocols, we first observe that we need
only show Alice and Bob can generate states close to $n$ EPR pairs.
This is a consequence of the following lemma (originally Note~28
of~\cite{qkd}):
 
\begin{lemma}
\label{thm:goodEPR}
If $\rho$ has a high fidelity $ 1 - 2^{-l}$ (for large $l$) to a state
of $n$ perfect EPR pairs and Alice and Bob measure along a common axis
to generate an $n$-bit key from $\rho$, then Alice and Bob will most
likely share the same key, which is essentially random. Moreover,
Eve's mutual information with the final key is bounded by $2^{-c} + O
(2^{-2l})$, where $c= l - \log_2 [ 2n +l + ( 1 / \log_e 2)]$.  In
other words, Eve's information is exponentially small as a function of
$l$.
\end{lemma}

The proof is given in Appendix~\ref{a:EPR}.  The next step is to
restrict our attention to Pauli attacks.

\begin{lemma} \cite{qkd}
\label{thm:Pauli}
Consider a stabilizer EPP protocol for QKD.
Given any joint attack $\A$ by Eve, there is a Pauli attack for which
the final density matrix $\rho_{AB}$ of Alice and Bob has the same
fidelity to $n$ EPR pairs, and which gives the same chance of having
the QKD protocol abort.
\end{lemma}
We will only prove Lemma~\ref{thm:Pauli}
for EPP protocols based on stabilizer EPPs,
but the result holds for any EPP designed to correct Pauli channels
(see~\cite{squeezed} for the general proof).
Pauli channels play a special role in the above
Lemma because most known quantum
error correcting codes (stabilizer codes, for instance)
are designed to correct Pauli errors.

\medskip \proof

First, note that for a symmetric EPP, it would suffice if Alice and
Bob had a way of measuring $M_\mu \otimes M_\mu$ directly instead of
separately measuring $M_\mu$ on Alice's side and again on Bob's side.
This is because all decisions are based on the parity of Alice's and
Bob's results, which is equal to the eigenvalue of $M_\mu \otimes
M_\mu$.  Also, note that the EPR pair $\ket{\Psi^+} = (\ket{00} +
\ket{11})/\sqrt{2}$ is a $+1$ eigenstate of the Pauli operators $X
\otimes X$ and $Z \otimes Z$.  (It is actually a $-1$ eigenstate of $Y
\otimes Y$.)

Thus, let $W_r$ be a Bell measurement for the $r$th EPR pair --- a
measurement of both $X \otimes X$ and $Z \otimes Z$.  For a stabilizer
EPP, $W_r$ {\it commutes} with $M_\mu \otimes M_\mu$ for all $\mu$, $r$
(note that each $M_\mu$ is likely to involve more than one EPR pair).
Thus, if Alice and Bob first measure all the operators $M_\mu \otimes
M_\mu$ and then measure $W_r$ for all $r$ after the EPP is concluded,
the result is the same as if they first measure $W_r$ and then $M_\mu
\otimes M_\mu$.  Since they do not need the results of the
measurements $W_r$, it is again equivalent if Eve measures $W_r$
instead of Alice and Bob.

That is, the following two situations are the same: a) Eve performs
her attack $\A$ and then Alice and Bob measure $M_\mu \otimes M_\mu$,
and b) Eve performs $\A$, measures $W_r$, and then Alice and Bob
measure $M_\mu \otimes M_\mu$.  By the argument of the previous
paragraph, the attack in b) produces a density matrix with the same
fidelity to n EPR pairs as the attack in a).  The attack $\A$ followed
by measurement of $W_r$ is a Pauli attack: The initial state is a Bell
state (the tensor product of $\ket{\Psi^+}$ for all pairs), and the
final state is a mixture of tensor products of Bell states (the
outcome of measuring $W_r$ for each pair $r$).  Each tensor product
$\ket{\Phi_j}$ of Bell states can be associated with the unique Pauli
operation $\Pl_j$ that maps $\ket{\Psi^+}^{\otimes N}$ to
$\ket{\Phi_j}$, so Eve's attack is $(\Pl_j, q_j)$, where $q_j$ is the
probability of getting the outcome $\ket{\Phi_j}$.  Therefore, the
lemma holds for a hypothetical protocol in which Alice and Bob measure
$M_\mu \otimes M_\mu$ directly.

Of course, Alice and Bob have no way of doing this, so instead they
must measure $M_\mu$ separately and compare results (with one- or
two-way communications, as appropriate).  Since this gives them more
information, it certainly cannot help Eve.  On the other hand, they
don't actually use that information --- from the definition of a
symmetric EPP, only the relative measurement outcome between Alice and
Bob matters. Therefore, having Alice and Bob measure $M_\mu \otimes
M_\mu$ together produces the same fidelity and chance of aborting as
when they measure $M_\mu$ separately.  \qed

This Lemma is described in~\cite{qkd} as a ``classicalization'' or
``quantum-to-classical reduction'' because it reduces Eve's general
quantum attack to a Pauli attack, which is classical in the sense that
it can be described by classical probability
theory. Lemma~\ref{thm:Pauli} allows us to simplify our discussion to
just Pauli channels $(\Pl_i, q_i)$.

We can simplify further by taking into account the symmetry of the QKD
protocol.  Note that in the EPP protocols we described, Alice and Bob
permute their qubits randomly before doing any other operations. So we
may as well assume $q_i = q_j$ whenever $\Pl_i$ is a permutation of
$\Pl_j$. That is, the attack is symmetric on the EPR pairs.
Similarly, in the two-basis scheme, Alice performs randomly one of the
two operations $I$, $H$, which produces a symmetry between the $X$ and
$Z$ bases, so we can also assume $q_i = q_j$ whenever $\Pl_i$ is
related to $\Pl_j$ by the Hadamard transform on any number of qubits.
In the three-basis scheme, we can assume $q_i = q_j$ when $\Pl_i$ and
$\Pl_j$ are related by $T$ or $T^2$ on some set of qubits.

Now, in the EPP protocols, Alice and Bob measure a random subset of
$m$ qubits to test the error rate.  From this, they are supposed to
figure out what sort of Pauli channel the system has undergone.  If
the noise occurs independently on each qubit, this is just a
straightforward problem in statistical inference.  Of course, an
eavesdropper need not use such a simple attack, but the symmetries of
the protocol still allow Alice and Bob to make a good guess as to the
true channel.  For one thing, Eve has no way to distinguish between
the test bits and the key bits, so the error rate measured for the
test bits should be representative of the error rate on the key bits.
What's more, Alice and Bob learn a good deal about the
basis-dependence of the channel as well.

Let us first consider the two-basis case more carefully.  Suppose
Alice and Bob find there are $p_I m_I$ errors among the $m_I$ qubits
for which Alice did the operation $I$; these represent $X$ and $Y$
Pauli errors introduced by Eve.  Similarly, they find $p_H m_H$ errors
in the $m_H$ qubits for which Alice did the operation $H$; these
represent $Y$ and $Z$ errors introduced by Eve.  If this channel were
an uncorrelated Pauli channel $(q^0_X, q^0_Y, q^0_Z)$, on average, we
would expect $p_I = q^0_X + q^0_Y$ and $p_H = q^0_Y + q^0_Z$.  In
fact, if we consider the effective error rates after undoing the $I$,
$H$ operations, we find $q_X = (q^0_X + q^0_Z)/2$ and $q_Z = (q^0_Z +
q^0_X)/2$ because $I$ and $H$ are equally likely.  That is, $q_X =
q_Z$.  The effective $Y$ error rate $q_Y = q^0_Y$.

Note that in the two-basis case, Alice and Bob are unable to deduce
the most likely values of $q_X$, $q_Y$, and $q_Z$; they can only learn
$p_X = q_Y + q_Z$ and $p_Z = q_X + q_Y$.  Given the symmetry between
$I$ and $H$, they in fact have $p_X = p_Z = (p_I + p_H)/2$, but our
discussion will keep $p_X$ and $p_Z$ as separate parameters.  This
allows most of our results to also apply to the efficient
case~\cite{eff,patent}, where $I$ and $H$ have different
probabilities.

The fact that Alice and Bob cannot completely learn the
characteristics of even an uncorrelated Pauli channel suggests that it
might be helpful to measure in more bases.  This is the advantage of
the six-state scheme, which is related to the three-basis EPP
protocol.  In that case, Alice and Bob measure $p_I$, $p_T$,
$p_{T^2}$.  For an uncorrelated Pauli channel $(q^0_X, q^0_Y, q^0_Z)$,
$p_I = q^0_X + q^0_Y$, $p_T = q^0_Y + q^0_Z$, and $p_{T^2} = q^0_X +
q^0_Z$.  Given the symmetry of the problem, after undoing the
rotations, we get $q_X = q_Y = q_Z = (q^0_X + q^0_Y + q^0_Z)/3 = (p_I
+ p_T + p_{T^2})/6$.  Again, our discussion will allow $q_X$, $q_Y$,
and $q_Z$ to be different to accommodate the efficient six-state
protocol.

Given the error test, Alice and Bob deduce some values either for
$p_X$, $p_Z$ or for all three quantities $q_X$, $q_Y$, $q_Z$.
However, the error rate on the tested bits is only {\em close} to the
error rate on the data bits.  Therefore, they should use an EPP that
is flexible enough to correct slightly more or less noisy Pauli
channels than indicated by the test.  In particular, when they deduce
$q_X$, $q_Y$, and $q_Z$, they should perform an EPP capable of
correcting any Pauli channel $(q_X^t, q_Y^t, q_Z^t)$ with $|q_i^t -
q_i| < \epsilon$ for $i = X, Y, Z$ and some small $\epsilon$.
Further, we should assume that, for any $\epsilon$, the fidelity of
the final state to $n$ EPR pairs is exponentially close to $1$ in $N$.

When Alice and Bob only learn $p_X$ and $p_Z$, they should allow
additional flexibility for the value of $q_Y^t$.  That is, their EPP
should correct any Pauli channel $(p_Z^t - a, a, p_X^t - a)$ (with all
three parameters non-negative), again with $|p_i^t - p_i| < 2\epsilon$,
for $i = X, Z$.  Provided Alice and Bob use such an EPP, the next
lemma says that the error test works and allows them to correct any
symmetric Pauli channel, not just an uncorrelated one.

\begin{lemma}
\label{thm:uncorrelated}
Suppose the $N$ EPR pairs experience a Pauli channel $(\Pl_i, q_i)$
which is symmetric over the $N$ pairs, and that they use an EPP which
corrects for any error rate close to those shown by the test bits, as
described above.  Then either they abort with high probability, or the
final state has fidelity exponentially close to $1$ in $N$ to the
state of $n$ EPR pairs.
\end{lemma}

Since we only need to consider Pauli channels, the proof is just an
exercise in classical probability, and is given in
Appendix~\ref{a:uncorrelated}.

From lemmas~\ref{thm:Pauli} and~\ref{thm:uncorrelated}, we know that
for the EPP protocols we consider, given any strategy for Eve, either
she has a large chance of getting caught, or the final state will have
high fidelity to $n$ EPR pairs.  Combining that with
lemma~\ref{thm:goodEPR}, we have shown:

\begin{thm}
\label{thm:secureEPP}
The EPP protocols for QKD are secure and correct.
\end{thm}

\subsection{Prepare-and-Measure Protocols are Secure}

Given theorem~\ref{thm:secureEPP}, Shor and Preskill~\cite{shorpre}
showed that one can prove the security of BB84.  The same technique
can be applied to show the security of the six-state
scheme~\cite{six}.  These two results can be combined into the
following theorem:

\begin{thm}[\cite{shorpre}]
\label{thm:reduction}
Given a QKD protocol based on a CSS-like 1-EPP, there exists a
``prepare-and-measure'' QKD protocol with similar security.  That is,
for any strategy by Eve to attack the ``prepare-and-measure''
protocol, there exists a strategy to attack the EPP protocol with
similar probability of causing the protocol to abort and similar
information gain to Eve if it does not abort.  (Similar here means
that the security parameters are polynomially related.)
\end{thm}

\proof

The reduction to a ``prepare-and-measure'' protocol is done as a
series of modifications to the EPP protocol to produce equivalent
protocols.  The main insight is that the $X$-type measurements do not
actually affect the final QKD protocol, and therefore are not needed.
The $X$-type measurements give the error syndrome for phase ($Z$)
errors, which do not affect the value of the final key.  Instead, $Z$
errors represent information Eve has gained about the key.  The phase
information thus must be delocalized, but need not actually be
corrected.  The upshot is that Alice and Bob need not actually measure
the $X$-type operators and can therefore manage without a quantum
computer.  Our initial goal is to manipulate the EPP protocol to make
this clear.  The $X$-type measurements do not, however, disappear
completely: instead they become privacy amplification.

For the first step, we modify the EPP to put it in a standard form.
Because it is a CSS-like 1-EPP, there is no branching in the tree
diagram, and each operator being measured is either $X$-type or
$Z$-type.  The operators all commute, and do not depend on the outcome
of earlier measurements, so we can reorder them to put all of the
$Z$-type measurements before all of the $X$-type measurements. 
Let us recall Definition~4 for a CSS-like 1-EPP. Now we
have an EPP consisting of a series of $Z$-type measurements, followed
by a series of $X$-type measurements, followed by CNOTs and Pauli
operations (which we can represent as $I$, $X$, and/or $Z$ on each
qubit).  Then Alice and Bob measure all qubits in the $Z$ basis.

As a second step, we can move all $X$ Pauli operations to before the
$X$-type measurements, since they commute with each other.  Moreover,
if Alice and Bob are simply going to measure a qubit in the $Z$ basis,
there is no point in first performing a $Z$ phase-shift operation,
since it will not affect at all the distribution of outcomes of the
measurement.

We now have an EPP protocol consisting of $Z$-type measurements,
followed by $X$ Pauli gates, followed by $X$-type measurements,
followed by a sequence of CNOT gates which does not depend on the
measurement outcomes.  But nothing in the current version of the
protocol depends on the outcomes of the $X$-type measurements, so
those measurements are useless.  We might as well drop them.
Furthermore, $X$ Pauli operations and CNOT gates are just classical
operations, so we might as well wait to do them until after the $Z$
basis measurement, which converts the qubits into classical bits.

What's more, it is redundant to perform $Z$-type measurements followed
by measurement of $Z$ for each qubit.  We can deduce with complete
accuracy the outcome of each $Z$-type measurement from the outcomes of
the measurements on individual qubits.  For instance, if a sequence of
three bits is measured to have the value $101$, then we know that
measurement of $Z_1 \otimes Z_2 \otimes Z_3$ will give the result $+1$,
as the parity of the three bits is even.

Thus, we are left with the following protocol: Alice prepares a number
of EPR pairs, and sends half of each to Bob.  She and Bob each perform
the correction rotation ($I$ or $H$ for the two-basis scheme, $I$,
$T$, or $T^2$ for the three-basis scheme), then measure each qubit in
the $Z$ basis.  They use some of the results to test the error rate,
and on the rest they perform some classical gates derived from the
original EPP.  

In fact, since Alice can perform her rotation and measurement before
sending any qubits to Bob, she need not actually prepare entangled
states.  Instead, she simply generates a random number, which
corresponds to the outcome of her $Z$ basis measurement, and sends Bob
the state to which the EPR pair would have collapsed, given that
measurement result.  That is, she sends him $\ket{0}$ or $\ket{1}$
rotated by the appropriate gate ($I$, $H$, $T$, or $T^2$).  Bob
inverts the rotation and measures.

Then they perform classical gates.  To understand which gates, it is
helpful to look more closely at the original EPP.  When the EPP is
based on a CSS code, the $Z$-type operators correspond to the parity
checks of a classical error-correcting code $C_1$, and the $X$-type
operators correspond to the parity checks of another classical code
$C_2$, with $C_2^\perp \subseteq C_1$.  The quantum codewords of the
CSS code are superpositions of all classical codewords from the cosets
of $C_2^\perp$ in $C_1$.  Measuring the $Z$-type operators therefore
corresponds to determining the error syndrome for $C_1$, whereas
measuring the $X$-type operators determines the error syndrome for
$C_2$.  The usual 1-EPP protocol for correcting errors is for Bob to
compute the difference, in both bases, between Alice's syndrome and
his syndrome, and then to perform a Pauli operation to give his state
the same syndromes as Alice's state.  That is, Alice and Bob now each
have a superposition over the same coset of $C_2^\perp$ within the
same coset of $C_1$ (or rather, they have an entangled state, a
superposition over all possible shared cosets for a given pair of
syndromes).  The decoding procedure then determines {\em which} coset
of $C_2^\perp$ they share and uses that as the final decoded state.

More concretely, we can describe the classical procedure as follows:
For the error correction stage, Alice computes and announces the
parity checks for the code $C_1$.  Bob subtracts his error syndrome
from Alice's and flips bits (according to the optimal error correction
procedure) to produce a state with $0$ relative error syndrome; that
is, he should now have the same string as Alice.  Then Alice and Bob
perform privacy amplification: they compute the parity checks of
$C_2^\perp$ (i.e., they multiply by the {\em generator} matrix of
$C_2$) and use those as their final secret key bits.

There is one final step to convert the protocol to a
``prepare-and-measure'' protocol.  Instead of preparing $N$ qubits and
sending them to Bob, Alice prepares $2N (1+\epsilon)$ (for BB84) or
$3N (1+\epsilon)$ (for the six-state scheme).  And instead of waiting
for Alice to announce which rotation she has performed ($I$, $H$, $T$,
or $T^2$), Bob simply chooses one at random.  Instead of rotating and
then measuring in the $Z$ basis, Bob simply measures in the $X$, $Y$,
or $Z$ basis, depending on which rotation he chose.  Then Alice and
Bob announce their bases, and discard those bits for which they
measured different bases.  With high probability, there will be at
least $N$ remaining bits.  Alice and Bob perform the error test on $m$
of them, and do error correction and privacy amplification on the
remaining $N-m$.  Since the discarded bits are just meaningless noise,
they do not affect the security of the resulting
``prepare-and-measure'' protocol.  The only difference is that
security must now be measured in terms of the remaining bits rather
than the original number of qubits sent.  When we begin with a
two-basis scheme, we end up with BB84; when we begin with a
three-basis scheme, we end up with the six-state
protocol.  \qed

\section{Difficulty in generalization to two-way EPPs}
\label{s:naive}

An obvious attempt to generalize theorem~\ref{thm:reduction} to
two-way EPPs would be to simply use CSS-like (those
with $X$-type and $Z$-type measurement operators only) 2-EPPs instead of
CSS-like 1-EPPs.  Unfortunately, this approach fails; another
condition is needed.

For instance, consider the following two-way EPP, which we call EPP~1:
Alice and Bob each measure $Z \otimes Z$ on pairs of EPR pairs.  This
can be implemented as a bilateral XOR: Alice performs an XOR from the
first pair to the second, and Bob does the same. Then both Alice and
Bob measure their qubit in the second pair and broadcast the
measurement result.  If Alice's and Bob's measurement outcome
disagree, they discard both pairs. On the other hand, if Alice's and
Bob's measurement outcome agree, then they keep the first pair for
subsequent operations. Now, if there is exactly one bit flip error
between the two pairs, Alice and Bob will disagree; otherwise they
agree. Note that at most one EPR pair out of the original two would
survive the measurement, but if Alice and Bob disagree, they discard
both pairs.  They do this for a large number of pairs; the surviving
EPR pairs have a lower bit flip error rate than the original ones.

Unfortunately, the surviving pairs also have a {\em higher} rate of
phase errors, since phase errors propagate backwards along a CNOT.
Therefore, in the next round of the EPP, Alice and Bob measure $X
\otimes X$ on pairs of EPR pairs.  This can be implemented by
performing a Hadamard transform, followed by the bilateral XOR and
measurement described above.  Alice and Bob should then perform
another Hadamard to return the surviving EPR pair to its original
basis.  This procedure can detect the presence of a single phase error
in the two pairs.  If Alice and Bob discard EPR pairs for which their
measurement results disagree, the surviving pairs will have a lower
rate of phase errors than before.

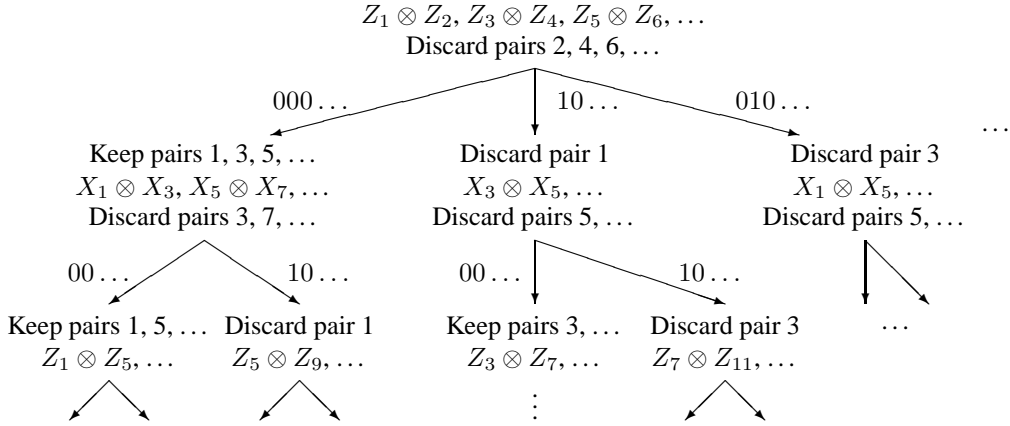
\begin{figure}

\centering

\begin{picture}(400,170)

\put(200,153){\makebox(0,0){$Z_1 \otimes Z_2$, $Z_3 \otimes Z_4$,
$Z_5 \otimes Z_6$, \ldots}}
\put(200,141){\makebox(0,0){Discard pairs 2, 4, 6, \ldots}}

\put(115,121){\makebox(0,0){$000\ldots$}}
\put(200,133){\vector(-4,-1){100}}
\put(75,100){\makebox(0,0){Keep pairs 1, 3, 5, \ldots}}
\put(75,88){\makebox(0,0){$X_1 \otimes X_3$, $X_5 \otimes X_7$,
\ldots}}
\put(75,76){\makebox(0,0){Discard pairs 3, 7, \ldots}}

\put(220,121){\makebox(0,0){$10\ldots$}}
\put(200,133){\vector(0,-1){25}}
\put(200,100){\makebox(0,0){Discard pair 1}}
\put(200,88){\makebox(0,0){$X_3 \otimes X_5$, \ldots}}
\put(200,76){\makebox(0,0){Discard pairs 5, \ldots}}

\put(290,121){\makebox(0,0){$010\ldots$}}
\put(200,133){\vector(4,-1){100}}
\put(325,100){\makebox(0,0){Discard pair 3}}
\put(325,88){\makebox(0,0){$X_1 \otimes X_5$, \ldots}}
\put(325,76){\makebox(0,0){Discard pairs 5, \ldots}}

\put(375,110){\makebox(0,0){\ldots}}

\put(35,56){\makebox(0,0){$00\ldots$}}
\put(75,68){\vector(-3,-2){36}}
\put(39,35){\makebox(0,0){Keep pairs 1, 5, \ldots}}
\put(39,23){\makebox(0,0){$Z_1 \otimes Z_5$, \ldots}}

\put(118,56){\makebox(0,0){$10\ldots$}}
\put(75,68){\vector(3,-2){36}}
\put(111,35){\makebox(0,0){Discard pair 1}}
\put(111,23){\makebox(0,0){$Z_5 \otimes Z_9$, \ldots}}

\put(183,56){\makebox(0,0){$00\ldots$}}
\put(200,68){\vector(0,-1){24}}
\put(200,35){\makebox(0,0){Keep pairs 3, \ldots}}
\put(200,23){\makebox(0,0){$Z_3 \otimes Z_7$, \ldots}}

\put(266,56){\makebox(0,0){$10\ldots$}}
\put(200,68){\vector(3,-1){72}}
\put(272,35){\makebox(0,0){Discard pair 3}}
\put(272,23){\makebox(0,0){$Z_7 \otimes Z_{11}$, \ldots}}

\put(325,68){\vector(0,-1){24}}
\put(325,68){\vector(1,-1){24}}
\put(337,35){\makebox(0,0){\ldots}}

\put(39,15){\vector(-1,-1){15}}
\put(39,15){\vector(1,-1){15}}

\put(111,15){\vector(-1,-1){15}}
\put(111,15){\vector(1,-1){15}}

\put(272,15){\vector(-1,-1){15}}
\put(272,15){\vector(1,-1){15}}

\put(200,8){\makebox(0,0){\vdots}}

\end{picture}

\caption{Tree diagram for EPP~1}
\label{fig:EPP1}
\end{figure}

The bit flip error rate has increased again.  However, the net effect
of the two rounds taken together has been to decrease both the $X$ and
$Z$ error rates (provided the error rates are not too high to begin
with).  Alice and Bob can continue to repeat this procedure, measuring
$X \otimes X$ alternately with $Z \otimes Z$, and the error rates will
continue to improve.  However, each round reduces the population of
EPR pairs by at least half, so a better strategy is to switch to a
more efficient one-way EPP once the error rates have dropped to the
point where one is viable.  Provided the initial error rate is not too
large, this procedure eventually converges.  The tree diagram for
EPP~1 is given in figure~\ref{fig:EPP1}.

The whole procedure only consists of measuring operators which are
either $X$-type or $Z$-type, so the EPP is CSS-like.  Still, we cannot
convert this EPP to a ``prepare-and-measure'' BB84 QKD scheme.

What goes wrong?  As is clear from figure~\ref{fig:EPP1}, the EPP
described is very definitely a two-way EPP, not a one-way EPP.  In
order to know which measurement to perform for the second round of the
protocol, both Alice and Bob must know which EPR pairs survived the
first round.  Similarly, in the third round, they must know which EPR
pairs survived the second round, and so forth.

In a ``prepare-and-measure'' scheme, Alice and Bob make all their
measurements in the $Z$ basis, and ignore the $X$-basis parity checks
because phase errors have no direct effect on the final key.  They can
therefore easily deduce the values of any operators which are the
product of all $Z$'s, but have no way of figuring out the measurement
result for a product of all $X$'s.  Since the second round consists of
measuring $X$ operators, Alice and Bob have no way of determining
which bits to keep for the third round of the protocol, and therefore
cannot complete the third round of the error correction/detection
process.  That is, they do not know along which branch in the tree
diagram they should proceed.

In a more intuitive language, the problem is that Alice and Bob do not
have quantum computers in a prepare-and-measure protocol. Therefore,
they cannot compute the phase error syndrome, which corresponds to the
eigenvalues of the $X$-type operators. For this reason,
they do not know which photons to throw away (conditional on the
phase error syndrome) and cannot complete the QKD process.

\section{Two-Way QKD}
\label{secmain}

Having understood the failure of EPP~1, we now present a generalization of
theorem~\ref{thm:reduction}.

\begin{thm}[Main Theorem]
\label{main}
Suppose a two-way EPP is CSS-like and also satisfies the following
conditions:
\begin{enumerate}
\item The tree diagram only branches at $Z$-type operators, not at
$X$-type operators.
\item The final decoding operations $U_{\mu}$ can depend arbitrarily
on the outcome of the measured $Z$-type operators, but cannot
depend on the outcomes of
the measured $X$-type operators at all.
The correction operation $P_{\mu}$ can depend on the outcome of
$X$-type operators, but only by factors of $Z$.
\end{enumerate}
Then the protocol can be converted to a ``prepare-and-measure'' QKD
scheme with security similar to the EPP-based QKD scheme.
\end{thm}

To understand these conditions, recall that the outcomes of $X$-type
operators represent the phase error syndrome.  Taken together, the two
conditions say that the outcomes of $X$-type operators are used to
perform phase error correction (by the factors of $Z$ in the
correction operator $P_{\mu}$), but nothing else.  For instance, no
post-selection based on the phase error syndrome is allowed.  From
there, the intuition is identical to that for the proof of the
Shor-Preskill result (theorem~\ref{thm:reduction}): Phase errors do
not affect the value of the key, so there is no need for Alice and Bob
to compute the phase error syndrome at all.  Therefore, Alice and Bob
do not really need quantum computers and can execute a
``prepare-and-measure protocol'' instead.  

\begin{figure}

\centering

\begin{picture}(350,130)

\put(200,112){\makebox(0,0){$Z$-type}}

\put(200,104){\vector(-4,-1){100}}
\put(100,71){\makebox(0,0){$Z$-type}}

\put(200,104){\vector(4,-1){100}}
\put(300,71){\makebox(0,0){$X$-type}}

\put(100,63){\vector(-2,-1){50}}
\put(50,30){\makebox(0,0){$X$-type}}

\put(100,63){\vector(2,-1){50}}
\put(150,30){\makebox(0,0){$Z$-type}}

\put(300,63){\vector(0,-1){25}}
\put(300,30){\makebox(0,0){$Z$-type}}

\put(50,22){\vector(0,-1){20}}

\put(150,22){\vector(-1,-1){20}}
\put(150,22){\vector(1,-1){20}}

\put(300,22){\vector(-1,-1){20}}
\put(300,22){\vector(1,-1){20}}

\end{picture}

\caption{Tree diagram of a 2-EPP satisfying the conditions of
theorem~\ref{main}.}
\label{fig:mainEPP}
\end{figure}
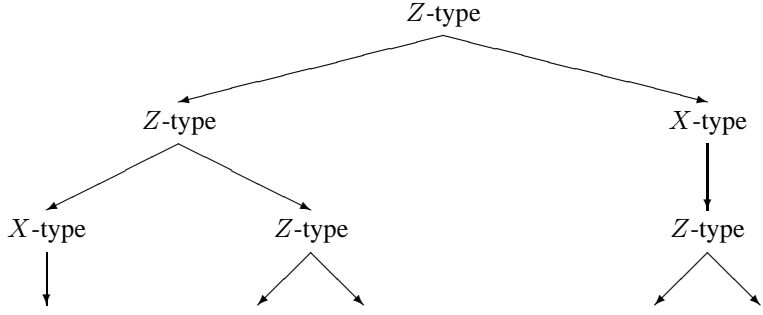

The tree diagram of a 2-EPP satisfying the conditions of this theorem
might look like the one depicted in figure~\ref{fig:mainEPP}.
The ``prepare-and-measure'' protocol produced by this theorem has the
following form:

\begin{enumerate}

\item Alice sends Bob $2N (1+\epsilon)$ qubits, randomly choosing 
$|0\rangle$ or $|1\rangle$ for each and putting each in either the $X$
or $Z$ basis at random.

\item Bob chooses to measure each qubit in the $X$ or $Z$ basis at
random.

\item Alice and Bob compare their measurement bases and discard those
qubits for which the bases disagree.  They keep $N$ remaining qubits.

\item Alice and Bob use $m$ of the qubits to estimate the error rate
from the channel, getting values $p_X$ and $p_Z$.

\item They now perform a combination of classical two-way error
detection/correction and classical privacy amplification based on the EPP.
The outcomes of $Z$'s serve two different functions:
``advantage distillation'' and also error correction.  
Indeed, Alice and Bob's ability to choose which branch to follow
(e.g., which EPR pairs to keep or throw away)
depending on the $Z$ operators means that Alice and Bob can
perform error detection. Not necessarily all bit-flip errors
are corrected. Since this is highly analogous to
the ``advantage distillation'' procedure in classical cryptography,
we will use the same name to denote such a procedure.
In addition, the $Z$
operators measured in the EPP can also act as classical parity checks
performed for error correction.
Finally, the $X$ operators measured become parities
extracted for privacy amplification.  If $M_\mu$ is an $X$-type operator, let
$v_\mu$ be a vector which is $1$ for any coordinate where $M_\mu$ has an
$X$, and is $0$ for any coordinate where $M_\mu$ acts as the identity
$I$. Consider the vector space $V$ generated by the $v_\mu$'s 
for consecutive $X$-type operators.
Then extract the parity for all vectors $u$ in the dual space,
$V^\perp$, of $V$.
These become the bits used in the next
round of error correction.

\end{enumerate}

\section{Another Two-Way EPP}
\label{EPPB}

Before proving the main theorem, we give an example of a two-way EPP
that satisfies the conditions of the theorem.  Like EPP~1, it will
consist of alternating rounds of measurements designed to handle bit
flip errors (``B steps'') and phase errors (``P steps'').

\medskip \noindent 
{\bf B step}: A B step is just the same as the first round of EPP~1:
Randomly permute all the EPR pairs. Afterwards,
perform a bilateral XOR between pairs of EPR pairs, and measure one of
the output pairs.  This effectively measures the operator $Z \otimes
Z$ for each of Alice and Bob, and detects the presence of a single bit
flip error.  Again, if Alice and Bob's measurement outcomes disagree,
they discard the remaining EPR pair.

\medskip
Note that this is similar to a classical protocol by Maurer
for advantage distillation~\cite{maurer0}.

The second round must deal with phase errors; however, we will not be
able to discard EPR pairs based on the result, since the conditions of
the theorem bar us from altering our protocol based on the measurement
results.  Instead, we take inspiration from the classical repetition
code.

A simple way to correct a single bit flip error is to use the majority
vote and encode the state $| 0 \rangle \mapsto | 000 \rangle,\ | 1
\rangle \mapsto | 111 \rangle$.  Therefore,
\begin{equation}
\alpha | 0 \rangle + \beta
| 1 \rangle \mapsto \alpha | 000 \rangle + \beta | 111 \rangle.
\end{equation}
Suppose the system is now corrupted by some bit flip errors.  A single
bit flip error can be detected by performing a majority vote. More
precisely, one measures $Z_1 Z_2$ to see if the first bit agrees with
the second bit and also $Z_1 Z_3$ to see if the first bit agrees with
the third bit.  These two measurements can be done coherently.  The
outcomes of the measurements are collectively called the error
syndrome and can be used to correct the state coherently.

The three-qubit bit flip error correction procedure can be turned into
a three-qubit phase error correction procedure by simply applying the
Hadamard transform, and into an EPP, following BDSW~\cite{BDSW}.

\medskip \noindent 
{\bf P step}: Randomly permute all the EPR pairs.
Afterwards, group the EPR pairs into sets of three, and measure $X_1
X_2$ and $X_1 X_3$ on each set (for both Alice and Bob). This can be
done (for instance) by performing a Hadamard transform, two bilateral
XORs, measurement of the last two EPR pairs, and a final Hadamard
transform.  If Alice and Bob disagree on one measurement, Bob
concludes the phase error was probably on one of the EPR pairs which
was measured and does nothing; if both measurements disagree for Alice
and Bob, Bob assumes the phase error was on the surviving EPR pair and
corrects it by performing a $Z$ operation.

\medskip
When there is only a single phase error among the three EPR pairs,
this procedure outputs a single EPR pair with no phase error.
However, when there are two or three phase errors, the final EPR pair
always has a phase error.  Therefore, when the phase error rate is low
enough, iteration of this procedure will improve it indefinitely,
while for higher phase error rates, the state will actually get worse.

The complete EPP protocol (EPP~2) consists of alternating B and P
steps for a number of rounds, until the effective error rate has
decreased to the point where one-way EPPs can take over.  Then we
decide on an appropriate CSS code and perform the corresponding
one-way EPP.  To get optimal performance, we should in fact use {\it
asymmetric} CSS codes, which correct a fraction $f_1$ of bit-flips and
a different fraction $f_2$ of phase errors. Note that, whenever $ 1 -
H (f_1) - H(f_2) \geq 0$, asymptotically, an asymmetric CSS code
exists that will correct those fractions of errors with high fidelity.
(A better bound might be obtained by considering the correlations
between bit-flip and phase errors. See \cite{six} for details.)  We
can view the whole EPP protocol as a kind of two-way concatenated
code.

EPP~2 satisfies the conditions of theorem~\ref{main}: it is CSS-like,
and measurements do not branch based on the outcome of $X$-type
measurements (which only occur during P steps and in the final CSS
code).  Furthermore, we only do Pauli operations based on the outcome
of $X$-type measurements.  Thus, we can apply theorem~\ref{main} to
convert EPP~2 into the following ``prepare-and-measure'' QKD scheme:

\bigskip
\begin{centering}
{\bf Protocol 2: repeated concatenation of BXOR with the three-qubit
phase code}
\end{centering}

\begin{enumerate}

\item Alice sends Bob a sequence of $N$ single photons as in either
BB84 or the six-state scheme.

\item Alice and Bob sacrifice $m$ of those pairs to perform the
refined data analysis.  They abort if the error rates are too large.

\item Alice and Bob randomly pair up their photons.  Alice publicly
announces the parity (XOR) of the bit values of each pair of her
photons, say $x_{ 2i-1} \oplus x_{2i}.$ Bob publicly announces the
parity (XOR) of his corresponding pair of photons, say $y_{2i-1}
\oplus y_{2i}$. If their parities agree, they keep one of the bits
from the pair --- i.e., Alice keeps $x_{2i-1}$ and Bob keeps
$y_{2i-1}$. If their parities disagree, they throw away the whole
pair.  (This step comes from a B step.)

\item Alice and Bob randomly form trios of the remaining bits and
compute the parity of each trio. They now regard those parities as
their effective new bits.  (This step comes from a P step.)

\item Steps 3) and 4) are repeated a prescribed number of times, say
$r$, which depends on the error rate measured in step 2.

\item Alice and Bob randomly permute their pairs.
They then apply a modified Shor and Preskill error
correction/privacy amplification procedure.  That is, Alice randomly
picks a codeword $u$ in the code $C_1$ and broadcasts $u+w$ to Bob,
where $w$ is her remaining bit string. Owing to the remaining noise in
the channel, Bob's current bit string is instead $w+e$.  He now adds
$u+w$ to his string to obtain a corrupted string $u+e$. He can apply
error correction for the code $C_1$ to recover $u$.  Here we use a
modified Shor and Preskill procedure that is based on an asymmetric
CSS code that corrects up to a fraction, $f_1$, of bit-flip errors and
a different fraction, $f_2$, of phase errors.

\item Alice and Bob perform the coset extraction procedure to
obtain the coset $u + C_2$, which gives their final key.

\end{enumerate}

In order to determine if the resulting QKD protocol is secure or not
at a given error rate, we need only study the behavior of EPP~2.
Furthermore, by lemmas~\ref{thm:Pauli} and~\ref{thm:uncorrelated} and
the intervening discussion, we need only study the behavior of EPP~2
for uncorrelated Pauli channels with nice symmetry properties.

For the six-state scheme, this is completely straightforward: we just
plug in the upper bounds on the error rates $(q_X, q_Y, q_Z)$ and see
if EPP~2 converges. This upper bound on the error rates gives the
worst case behavior. For the usual six-state scheme, we may even
assume $q_X = q_Y = q_Z = q$.  We can test for convergence with a
simple computer program; we follow the error rates through B and P
steps until they are small enough so that CSS coding is effective.  If
the program indicates convergence for $q$, the EPP definitely
converges, and we have proved the six-state protocol is secure at bit
error rate $q$.  In this way, we have shown the six-state scheme
remains secure to an error rate of at least $23.6\%$.  If the program
does not converge, that does not necessarily imply that the six-state
scheme is insecure using this post-processing method; it simply means
it did not converge within the regime where our program is numerically
reliable.

A study of BB84 is slightly more difficult.  Alice and Bob do not know
$(q_X, q_Y, q_Z)$, only $p_X = q_Y + q_Z$ and $p_Z = q_X + q_Y$.
There is one free parameter $q_Y = a$; then, for BB84, $q_X = q_Z = p
- a$, where $p=p_X=p_Z$ is the bit error rate.  To show that BB84 is
secure using this post-processing scheme, we must show that EPP~2
converges for all values of $a\in [ 0, p]$.  However, this is not
immediately
compatible with a numerical approach, since we would have to check
infinitely many values of $a$.  Instead, we first show analytically
that $a=0$ (no
$Y$ errors) gives the worst case; the proof is in
appendix~\ref{a:worst}.  Then we need only check in our
program that EPP~2 converges
for the uncorrelated Pauli channel $(p, 0, p)$.  Our program then
indicates that BB84 is secure to an error rate of at least $17.9\%$.

It turns out, however, that alternating B and P steps is not optimal.
EPPs based on other arrangements of these two steps can converge at
higher error rates.  For instance, for the three-basis protocol, we
have discovered that a sequence of five B steps, followed by
asymmetric CSS coding, converges to an error rate of at least
$26.4\%$, and that therefore the six-state scheme remains secure to at
least this bit error rate.  Similarly, setting $a=0$ in the two-basis
protocol, a sequence of five B steps, followed by six P steps,
followed by asymmetric CSS coding converges up to an error rate of at
least $18.9\%$.  Since $a=0$ is again the worst case, this shows that
BB84 can be secure to at least this bit error rate.

We remark that, in the above discussion, we have assumed that Alice
and Bob simply throw away the error syndrome of each round immediately
after its completion. Such an assumption greatly simplifies our
analysis. However, in principle, Alice and Bob can employ an improved
decoding scheme where they keep track of all the error syndromes and
use them to improve the decoding in future rounds of the algorithm.
It would be interesting to investigate in the future how much the
tolerable error rates can be increased by such an improved decoding
scheme.  Of course, other improvements might be possible as well,
including different kinds of B and P steps.
The threshold error rate (i.e., the maximal bit error rate that
can be tolerated) of
a prepare-and-measure QKD scheme remains an important open question.

\section{Proof of the Main Theorem}
\label{s:proof}

To prove theorem~\ref{main}, we begin with a QKD protocol using the
two-way EPP directly.  The security of this protocol follows
immediately from theorem~\ref{thm:secureEPP}.  As in the proof of
theorem~\ref{thm:reduction}, we then rearrange the protocol into a
standard form in which it is clear that the $X$-type measurements are
unnecesarry.  From there, it is an easy step to a prepare-and-measure
protocol.

\begin{enumerate}

\item \label{Hadamard}
Alice prepares $N$ EPR pairs.  She performs a Hadamard transform
on the second qubit for half of them, chosen at random.

\item Alice sends the second qubit from each EPR pair to Bob.  Bob
acknowledges receiving them, and then Alice tells him which ones have
the Hadamard transform.  Bob reverses all Hadamard transforms.

\item Alice and Bob select $m$ EPR pairs to test the error rate in the
channel.

\item Alice and Bob perform the two-way EPP on the remaining $N-m$ EPR
pairs.  They now have a number of EPR pairs of very good fidelity.

\item Alice and Bob measure each remaining EPR pair in the $Z$ basis
to produce a secure shared key.

\end{enumerate}

The above protocol assumes a two-basis QKD scheme.  For a three-basis
scheme, Alice and Bob apply one of the three operations $I$, $T$,
$T^2$ instead of $I$ or $H$.

\begin{figure}
\begin{center}
\begin{picture}(260,80)

\put(20,60){\line(1,0){14}}
\put(46,60){\line(1,0){48}}
\put(106,60){\line(1,0){14}}
\put(20,40){\line(1,0){14}}
\put(46,40){\line(1,0){49}}
\put(20,20){\line(1,0){14}}
\put(46,20){\line(1,0){49}}

\put(34,54){\framebox(12,12){$H$}}
\put(34,34){\framebox(12,12){$H$}}
\put(34,14){\framebox(12,12){$H$}}

\put(60,60){\circle*{4}}
\put(60,60){\line(0,-1){24}}
\put(60,40){\circle{8}}

\put(80,60){\circle*{4}}
\put(80,60){\line(0,-1){44}}
\put(80,20){\circle{8}}

\put(94,54){\framebox(12,12){$H$}}

\put(105,34){\makebox(20,12){Meas.}}
\put(105,14){\makebox(20,12){Meas.}}

\put(136,34){\makebox(12,12){$=$}}

\put(160,60){\line(1,0){100}}
\put(160,40){\line(1,0){54}}
\put(226,40){\line(1,0){4}}
\put(160,20){\line(1,0){54}}
\put(226,20){\line(1,0){4}}

\put(180,40){\circle*{4}}
\put(180,40){\line(0,1){24}}
\put(180,60){\circle{8}}

\put(200,20){\circle*{4}}
\put(200,20){\line(0,1){44}}
\put(200,60){\circle{8}}

\put(214,34){\framebox(12,12){$H$}}
\put(214,14){\framebox(12,12){$H$}}

\put(240,34){\makebox(20,12){Meas.}}
\put(240,14){\makebox(20,12){Meas.}}

\end{picture}

\caption{Two equivalent ways to measure the operators $X_1 X_2$ and
$X_1 X_3$.}
\label{fig:Pstep}
\end{center}
\end{figure}
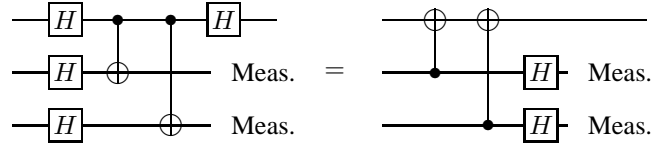

To reduce the above EPP protocol to a prepare-and-measure one, we
would like to eliminate the phase error correction steps in the EPP
protocol.  For a CSS-like EPP, phase error correction comes completely
from the measurement of $X$-type operators $M_\mu$.  We can perform
such a measurement as a Hadamard transform, followed by a series of
CNOTs with the same target qubit.  Then we measure the target qubit,
and Hadamard transform the others back to the original basis (see, for
instance, the left network in fig.~\ref{fig:Pstep}).  This procedure
computes the parity of all the control qubits and the target qubit in
the $X$ basis, and gives the eigenvalue of $M_\mu$.  (Of course, in
the context of an EPP, each of Alice and Bob perform this procedure,
and compare results.)

However, this series of gates --- Hadamard, CNOT, Hadamard --- is
equivalent to a single CNOT gate with control and target reversed.
This means, for example, that the two circuits depicted in
fig.~\ref{fig:Pstep} are mathematically equivalent.
Note that the right hand side depicts an essentially {\em classical} 
circuit composed of CNOTs (with a couple of X-basis measurements at
the end).  Instead of working with a quantum
circuit for phase error correction, as depicted by the left hand side
of the figure, one can work with the essentially classical circuit in 
the right hand side.

The same principle holds in general.  $X$-basis measurements can
be written as effectively classical circuits consisting of
a series of CNOTs (with the same control qubit but different 
target qubits), followed by a Hadamard transform and measurement on 
the control qubit.  The qubits which survive the procedure have only
experienced the CNOT gates. So it will be easy to convert this circuit
to a truly classical one.

Note that each target qubit gets replaced 
by its XOR with the control qubit; in other words, by a parity which 
is orthogonal to the vector $v_\mu$
derived from $M_\mu$ by replacing $X$'s with $1$'s.
For instance, in our sample EPP~2, we measure two $X$ operators 
in a row for a set of three qubits, $X_1 X_2$ and $X_1 X_3$.  The 
effect of these measurements in the $Z$ basis is to map 
$|a,b,c\rangle \mapsto |a+b,b,c\rangle \mapsto |a+b+c,b,c\rangle$.  
That is, the first qubit gets replaced by the parity of all three 
qubits.  We could also see this by noting that the only nontrivial 
vector which is orthogonal to both $(1,1,0)$ and $(1,0,1)$ is $(1,1,1)$.

However, for EPPs satisfying the conditions of theorem~\ref{main}, no
choice of $M_\nu$ later in the protocol depends on the outcome of the
$X$-type measurement $M_\mu$.  Therefore, we can delay making the
actual measurement until the end of the protocol, after we have
measured all $Z$ operators.  The EPP may call for correcting phase
errors immediately by performing $Z$ rotations based on the
measurement results, but we can delay those as well using the
identities
\begin{eqnarray}
{\rm CNOT} (Z \otimes I) & = & (Z \otimes I) {\rm CNOT} \\
{\rm CNOT} (I \otimes Z) & = & (Z \otimes Z) {\rm CNOT}.
\end{eqnarray}
That is, we can move a $Z$ rotation from before a CNOT to after it,
possibly at the price of having to do two of them instead.
Ultimately, we end up with a circuit consisting only of $Z$-basis
measurements and quantum CNOT gates (whose position may depend on the
outcome of a $Z$ measurement), followed by $X$-basis measurements and
phase shifts.  This is an equivalent EPP to the one we began with.

In the QKD protocol, after performing the EPP, Alice and Bob measure
each surviving EPR pair in the $Z$ basis to produce a key.  But phase
shifts are irrelevant if we are immediately going to measure in the
$Z$ basis, so Alice and Bob need not actually perform them or the
$X$-basis measurements controlling them. 

Alice and Bob now have a completely classical circuit, followed by
measuring all the qubits in the $Z$ basis.  They get the same result
if they instead measure all the qubits first, and {\em then} perform
the classical circuit.  The circuit they have is exactly the error
correction and privacy amplification protocol described in
sec.~\ref{secmain} as coming from the EPP.  Note that any
communication from Bob to Alice occurs during the classical
circuit {\em after} the initial measurement.

To complete the transformation to a ``prepare-and-measure'' protocol,
we follow a few additional steps from Shor and Preskill.  Instead of
preparing a number of EPR pairs and measuring them, Alice can just
generate a random bit string, and send Bob the state he would have
gotten if she made the EPR pairs and got that measurement result.
That is, she sends Bob a series of $0$s and $1$s chosen at random, and
puts half of them in the $X$ basis (when in the EPP protocol she would
perform a Hadamard transform in step~\ref{Hadamard}), and puts half of
them in the $Z$ basis (when there would be no Hadamard in the EPP
protocol).  Bob receives them, waits for Alice to tell him the basis,
and then measures in that basis.

Of course, we can wait to decide on the EPP until after Bob receives
his states, so it is equally good if Bob guesses a basis for each
qubit and measures immediately.  Then when Alice tells Bob which basis
she used, they discard any bits where the bases disagree.  This gives
the final ``prepare-and-measure'' protocol.

To prove the security of a six-state protocol, one
uses three bases $X$, $Y$, and $Z$ in the appropriate place instead of
just the $X$ and $Z$ bases.  Otherwise, the proof is identical.

\section{Concluding Remarks}

We have proven the unconditional security of standard quantum key
distribution schemes including BB84 and the six-state scheme.  Our
proof allows Alice and Bob to employ two-way classical
communications. Compared to previous schemes, it has the advantage of
tolerating substantially higher bit error rates. Indeed, we have
shown that the BB84
scheme can be secure even at a bit error rate of 18.9\% and the
six-state scheme at 26.4\%. By tolerating such high bit error rates,
our result may extend the distance of QKD and increase the key
generation rate.  Our result is conceptually interesting because it
may spur progress in the study of two-way entanglement purification
protocols (EPPs).  We have introduced a new subclass of two-way
entanglement purification protocols (EPPs) and demonstrated that such
a subclass of protocols can be reduced to standard BB84 and the
six-state scheme.  Our results demonstrate clearly that two-way
classical communications can be used to enhance the secrecy capacity
of a QKD scheme and also show the six-state scheme can intrinsically
tolerate a higher bit error rate than BB84.

Our versions of the BB84 and six-state QKD schemes require two-way
classical communications between Alice and Bob in the post-processing
step of classical data (i.e., in the error correction and privacy
amplification stage).  This is not a bad thing in itself because {\it
any} protocol of BB84 (or six-state) requires two-way classical
communications anyway.  Indeed, in the basis comparision step, Alice
and Bob publicly announce their bases and throw away the polarization
data that are transmitted and received in different bases.  In order
for both Alice and Bob to know which polarization data to keep, it is
necessary to employ two-way classical communications.  Of course,
the ``one-way'' classical post-processing schemes require fewer rounds
of communication (and therefore less time) to complete, so there
appears to be a tradeoff between round-complexity of the protocol and
tolerable error rate.

Relating to earlier work on QKD, we remark that we have provided the
first examples of unconditionally secure schemes for advantage
distillation\cite{maurer0,maurer1,maurer2} in QKD.  Finally, two-way
entanglement purification techniques may provide a simple way to
understand other security proofs.  For instance, in
Appendix~\ref{a:inamori}, we provide a simple derivation of Inamori's
security proofs \cite{inamori,inamori6}.
For future work, it would be interesting to take into account the
effects of imperfections including faulty photon sources, lossy
channels, and photon dark counts\cite{ilm}.

\section{Acknowledgment}
\label{acknowledge}
We particularly thank Jayson Cohen for his generous help in Matlab
programming, and David DiVincenzo and Debbie Leung for
suggesting the tree diagram representation of an EPP.
Enlightening discussions and references on bounds on error rates in
prior art QKD schemes have been kindly provided by
Norbert L\"{u}tkenhaus and Nicolas Gisin.
We have also greatly benefitted from helpful
communications with Nabil Amer, Charles Bennett, Hitoshi
Inamori, Ueli M. Maurer, Renato Renner, John Smolin, Barbara Terhal and,
particularly, John Preskill and Peter Shor.
This research was conducted in part during the period
that D.~G. served as a Clay Long-Term CMI Prize Fellow
and during a visit of H.-K. Lo at Caltech.

\appendix

\section{Proof of Lemma~\ref{thm:goodEPR}}
\label{a:EPR}

The statements that Alice and Bob will most likely share the same key
and that the key is essentially random are clear.  We will focus on
proving the bound on Eve's information.  The proof of this crucial
part of Lemma~\ref{thm:goodEPR} follows from the following two claims,
which originally appeared in supplementary Note~II of \cite{qkd}.

\begin{claim}[High Fidelity implies low entropy] 
If $ \langle n {\rm\ singlets} | \rho | n {\rm\ singlets} \rangle 
> 1 - \delta$,
where $\delta \ll 1$, then von Neumann entropy
$S (\rho) < - ( 1 - \delta) \log_2 ( 1 - \delta) - \delta
\log_2 { \delta \over ( 2^{2R} -1)}$.
\end{claim}

\proof \ 
If $ \langle n {\rm\ singlets} | \rho | n {\rm\ singlets} \rangle 
> 1 - \delta$,
then the largest eigenvalues of the density matrix must be larger than
$1 - \delta$.   The entropy of $\rho$ is, therefore, bounded above by
that of $\rho_0 = diag ( 1 - \delta, { \delta \over ( 2^{2R} - 1)},
{ \delta \over ( 2^{2R} - 1)}, \cdots, { \delta \over ( 2^{2R} - 1)} )$.
That is, $\rho_0$ is diagonal with a large entry $ 1 - \delta$
and with the remaining probability $\delta$ equally distributed between
the remaining $2^{2R} - 1$ possibilities.
\qed

\begin{claim}[Entropy is a bound to mutual information]
Given any pure state $\phi_{CD}$ of a system consisting of
two subsystems $C$ and $D$, and any generalized measurements
$X$ and $Y$ on $C$ and $D$ respectively, the entropy of each subsystem
$S(\rho_C)$ (where
$\rho_C = {\rm Tr}_D | \phi_{CD} \rangle \langle \phi_{CD} | $) is an
upper bound to the amount of mutual information between $X$ and $Y$.
\end{claim}

\proof\ This is a corollary to Holevo's theorem \cite{holevo}. \qed

\section{Proof of Lemma~\ref{thm:uncorrelated}}
\label{a:uncorrelated}

We wish to show that, given any (not necessarily uncorrelated) Pauli
channel, our procedure of testing the error rate and then choosing an
appropriate code actually does correct the errors with high
probability.  The idea is that, because of the random permutation, the
EPP treats symmetrically all errors with a given breakdown into $X$,
$Y$, and $Z$ errors (the ``type'' of $P_j$).  The type of the true
error will be close to the estimated type.  We then show that the EPP
performs well for the likely types of error.

Since the channel is symmetric over all $N$ pairs, the pairs chosen
for error testing are a representative sample, and the number of
errors of any given kind in the sample will be close to the number of
errors of the same kind in the remaining pairs.  What we mean by the
``same kind'' bears a little explanation.  As discussed before the
statement of Lemma~\ref{thm:uncorrelated}, we only directly measure
the presence of two out of the three types of error, depending on
which operation ($I$, $H$, $T$, or $T^2$) we perform.  For instance,
when $I$ is performed, we measure the presence of only $X$ or $Y$
errors.  However, since Eve has no knowledge of which operation is
used for any particular qubit, the sample of test bits with a
particular operation gives a good estimate of the number of the
appropriate pair of errors in the remaining qubits of the sample.  For
instance, the fraction of errors among $I$ test qubits gives us a good
estimate of the number of qubits with either $X$ or $Y$ errors in
them.  Then the deduced rates of $X$, $Y$, and $Z$ errors (as
discussed before Lemma~\ref{thm:uncorrelated}) give a good estimate of
the actual error rates in the untested pairs.

For any particular instance of the protocol, the channel performs a
particular $N$-qubit Pauli operation $P_j$ (with probability $q_j$).
For any particular $j$, let $q_i^d$ be the deduced fraction of errors
of type $i$ ($i = X, Y, Z$) in the sample and let $q_i^u$ be the
actual fraction of errors of type $i$ in the untested pairs (``d'' for
``deduced'' and ``u'' for ``untested'').  Then for large $N$, with
high probability,
\begin{equation}
|q_i^d - q_i^u| < \epsilon.  
\label{eq:errclose}
\end{equation}
(That is, the deduced error rate is close to the true error rate.)
Naturally, $q_i^d$ and $q_i^u$ will depend on $j$, but we suppress
this dependence to simplify the notation.

Let us now restrict attention to one particular set of values for
$q_i^d$ and $q_i^u$ (which need not be equal, but which satisfy
condition~(\ref{eq:errclose})).  If the $q_i^d$ are large, Alice and
Bob will abort the protocol.  Otherwise, we wish to show that the EPP
used by Alice and Bob will correct most errors with these parameters.

To see this, we note that the EPP will correct the uncorrelated Pauli
channel $(q_X^u, q_Y^u, q_Z^u)$ on $N-m$ EPR pairs with high fidelity
$F$.  Suppose the EPP gives fidelity $F_j$ whenever the $N$-qubit
Pauli operation $P_j$ occurs (for a stabilizer EPP, $F_j$ will be
either $0$ or $1$).  Then
\begin{equation}
F = \sum_j p_j F_j,
\end{equation}
where $p_j$ is the probability of the Pauli operation $P_j$ for the
{\em uncorrelated} Pauli channel (not the true channel).  We can break
the sum over $j$ into two parts.  The first part will consist of the
set $S$ of $j$ for which $P_j$ contains exactly $n_X = q_X^u (N-m)$
$X$ errors, $n_Y = q_Y^u (N-m)$ $Y$ errors, and $n_Z = q_Z^u (N-m)$
$Z$ errors (the $n_i$ are integers by the definition of $q_i^u$).  The
second part consists of all other $j$.  Now, let $p$ be the
probability of any particular error in $S$, so
\begin{equation}
\sum_{j \notin S} p_j F_j \leq \sum_{j \notin S} p_j =
1 - \sum_{j \in S} p,
\end{equation}
so
\begin{eqnarray}
F & = & \sum_{j \in S} p_j F_j + \sum_{j \notin S} p_j F_j \\
 & \leq & p \sum_{j \in S} F_j + 1 - p |S| \\
 & = & 1 - p |S| (1 - \sum_{j \in S} F_j/|S|). \label{eq:Fbound}
\end{eqnarray}
But 
\begin{equation}
p = (q_X^u)^{n_X} (q_Y^u)^{n_Y} (q_Z^u)^{n_Z} (q_I^u)^{n_I},
\label{eq:pj}
\end{equation}
where $q_I^u = 1 - (q_X^u + q_Y^u + q_Z^u)$ is the probability of
identity operations, and $n_I = q_I^u (N-m)$ is the actual number of
identity operations.  $S$ contains $(N-m)! / (n_X! n_Y! n_Z! n_I!)$
elements, so using Stirling's approximation, we find
\begin{equation}
p |S| \approx (\frac{2\pi}{N-m})^{3/2} 
\frac{1}{\sqrt{q_X^u q_Y^u q_Z^u q_I^u}}.
\end{equation}
This is only polynomially small in $N-m$.  In order for $F$ to be
exponentially close to $1$ in equation~(\ref{eq:Fbound}), we therefore
require that $\sum_{j \in S} F_j$ be $[1-\exp(-O(N))]\,\, |S|$.

Now we can approximate the fidelity of the EPP for the general Pauli
channel $(\Pl_i, q_i)$.  We again write $F = \sum q_i F_i$ (with the
same $F_i$s, which only depend on the EPP, not the channel), and
recall that $q_i = q_j = q_{n_X,n_Y,n_Z}$ whenever $i$ and $j$ have
the same numbers $(n_X, n_Y, n_Z)$ of $X$, $Y$, and $Z$ errors.  That
means we can write
\begin{equation}
F = \sum_{n_X, n_Y, n_Z} q_{n_X,n_Y,n_Z} \sum_{i \in S_{n_X,n_Y,n_Z}}
F_i.
\end{equation}
But, except with exponentially small probability, the values $n_X$,
$n_Y$, $n_Z$ are within the allowed $\epsilon$-sized window for the
EPP, which we have shown means that $\sum_{i \in S} F_i = [1 -
\exp(-O(N))] |S|$.  Thus,
\begin{equation}
F = \sum_{n_X, n_Y, n_Z} q_{n_X,n_Y,n_Z} |S| [1 - \exp(-O(N))] -
\exp(-O(N)).
\end{equation}
Since
\begin{equation}
\sum_{n_X,n_Y,n_Z} q_{n_X,n_Y,n_Z} |S| = 1,
\end{equation}
it follows that the fidelity for the general Pauli channel is
exponentially close to $1$.
\qed

\section{Proof that $a=0$ is the worst case}
\label{a:worst}

In this section, we will show that it is sufficient to check the $a=0$ case
(with no $Y$ errors) when determining convergence of the 2-EPPs we
study for the BB84 protocol.

\begin{thm}
Suppose an EPP starts with a B step, followed by any series of B
and/or P steps, followed by asymmetric CSS coding.  Suppose $0 \leq p
< 1/4$.  If the EPP converges for the uncorrelated Pauli channel $(p,
0, p)$, then it will also converge for all uncorrelated Pauli channels
$(p-a, a, p-a)$, with $0 \leq a \leq p$.
\end{thm}

The initial condition $p < 1/4$ simply ensures that (for any value of
$a$) the state is more likely to be correct than incorrect, and will
be satisfied easily by all parameter sets we consider.  In fact, when
$p \geq 1/4$, an intercept-resend attack defeats BB84 (see
section~\ref{s:protocol}).

\proof

To do this, we will need to look at the behavior of the three error
rates as we perform steps of the protocol.  After each B or P step,
there is a new set of effective error rates on the pairs surviving the
round.

It is worth noting two things about protocols of the given form:
First, if the initial density matrix comes from a Pauli channel, then
the effective channel after any number of rounds will also be a Pauli
channel.  This is because all operations are from the Clifford group,
which preserves the Pauli group.  Second, if the initial channel
causes errors which are uncorrelated between EPR pairs, this property
will also be preserved after an arbitrary number of B and P rounds.
This is because both B and P rounds keep at most one of the pairs
which interact, so there is no opportunity to create correlations
between pairs which survive to the next round.  Therefore, we can
completely describe the effective error rates at any given point in
the protocol by a triplet $(q_X, q_Y, q_Z)$.

Suppose we start with error rates $(q_X, q_Y, q_Z)$ and perform a B
step.  Given any of the 16 possible configurations of errors, we can
deduce whether the remaining pair is discarded, and if not, whether it
has an error, and what kind of error it is.  The new error rates on
the surviving pairs are then $(q_X', q_Y', q_Z')$:
\begin{eqnarray}
q_X' & = & (q_X^2 + q_Y^2)/p_S, \\
q_Y' & = & 2 q_X q_Y / p_S, \\
q_Z' & = & 2 (1 - q_X - q_Y - q_Z) q_Z / p_S, \\
p_S & = & 1- 2 (q_X + q_Y)(1 - q_X - q_Y),
\end{eqnarray}
where $p_S$ is the probability that a pair will survive the check.

If we have error rates $(q_X, q_Y, q_Z)$ and perform a P step, we get
new error rates $(q_X', q_Y', q_Z')$:
\begin{eqnarray}
q_X' & = & 3 q_I^2 (q_X + q_Y) + 6 q_I q_X q_Z + 3 q_X^2 q_Y + q_X^3, \\
q_Y' & = & 6 q_I q_Y q_Z + 3 q_X (q_Y^2 + q_Z^2) + 3 q_Y q_Z^2 + q_Y^3, \\
q_Z' & = & 3 q_I (q_Y^2 + q_Z^2) + 6 q_X q_Y q_Z + 3 q_Y^2 q_Z +
q_Z^3, \\
q_I & = & 1 - q_X - q_Y - q_Z,
\end{eqnarray}
where $q_I$ is the initial probability of no error.

To prove the theorem, we change variables.  Instead of working
with $(q_X, q_Y, q_Z)$, we will work with $(p_Z, p_X, \Delta)$:
\begin{eqnarray}
p_Z & = & q_X + q_Y \\
p_X & = & q_Y + q_Z \\
\Delta & = & q_Z - q_Y = p_X - 2a.
\end{eqnarray}
As $a$ increases, $p_X$ and $p_Z$ stay the same, while $\Delta$
decreases.  We will show that the protocol behaves worse for larger
$\Delta$, so the worst case is $a=0$.

In the new variables, a B step maps the error rates from $(p_Z, p_X,
\Delta)$ to $(p_Z', p_X', \Delta')$:
\begin{eqnarray}
p_Z' & = & p_Z^2 / p_S,\\
p_X' & = & \left[p_X - p_X^2 + \Delta (1 - 2 p_Z - \Delta)\right]/p_S
\\
\Delta' & = & \left[p_X (1-2p_Z) + \Delta (1-2p_X) \right]/ p_S \\
p_S & = & 1 - 2 p_Z + 2 p_Z^2.
\end{eqnarray}
Since $p_X, p_Z < 1/2$ always in the regime of interest, $\Delta'$ is
increasing in $\Delta$, and $p_Z'$ never depends on $\Delta$ at all.
Provided $1- 2p_Z - 2\Delta > 0$, $p_X'$ also increases with
$\Delta$.  When this is true, $\Delta'$ and $p_X'$ also both increase
with $p_X$.

A P step takes the error rates from $(p_Z, p_X, \Delta)$ to $(p_Z',
p_X', \Delta')$ with the following relations:
\begin{eqnarray}
p_Z' & = & 3 p_Z (1-p_Z)^2 + p_Z^3, \\
p_X' & = & 3 p_X^2 (1-p_X) + p_X^3, \\
\Delta' & = & 3 \Delta^2 (1 - 2 p_Z - \Delta) + \Delta^3.
\end{eqnarray}
This time, $p_X'$ and $p_Z'$ only depend on $p_X$ and $p_Z$,
respectively, never on $\Delta$.  $p_X'$ increases with $p_X$.
$\Delta'$ only depends on $\Delta$ and $p_Z$, and increases with
$\Delta$ if two conditions --- $1 - 2p_Z - \Delta > 0$
and $\Delta \geq 0 $ --- are simultaneously satisfied.

\begin{claim}
The following inequalities hold:
\begin{enumerate}
\item At all points after the initial B step, $\Delta \geq 0$.
\item $1 - 2 p_Z - 2 \Delta > 0$ always. 
\end{enumerate}
\end{claim}

Note that when $p_X + p_Z < 1/2$, so that at least half the time there
is no error, it follows that $1 - 2 p_Z - 2 \Delta > 0$, since $\Delta
< p_X$.  However, it is not clear if the condition $p_X + p_Z < 1/2$
is preserved under the B and P steps.

From this claim, the theorem will follow: consider running the
protocol starting with error rates $(p_Z, p_X, \Delta) = (p, p, p)$ or
$(p, p, \Delta_0)$, with $\Delta_0 < p$.  Since the value of $p_Z$ at
any given time only depends on the previous value of $p_Z$, $p_Z$ will
always be equal in these two cases.  At any time, $p_X$ for the first
case will be greater than or equal to $p_X$ for the second case, and
$\Delta$ for the first case will be greater than or equal to $\Delta$
for the second case.  This is true by induction: it is true initially,
and at all steps, $p_X'$ and $\Delta'$ increase with $p_X$ and
$\Delta$ from the previous step.  Thus, the worst case is when
$\Delta = p$, which means $a=0$.

\medskip \proof\ (of claim)

Immediately after the initial B step, $\Delta' \geq 0$, because in
this step, $p_X = p_Z = p$ by the symmetry of BB84, and $\Delta \geq -
p$.  After subsequent B steps, $\Delta' \geq 0$ if $\Delta \geq 0$,
since $1 - 2 p_X$ and $1 - 2 p_Z$ are always positive.

After a P step, $\Delta' \geq 0$ if $3(1 - 2 p_Z - 2/3 \Delta) \geq
0$.  This will immediately follow if we can show $1 - 2 p_Z - 2 \Delta
> 0$, since before a P step, $\Delta \geq 0$ always.  Then by
induction, we will have shown $\Delta \geq 0$ at all points after the
initial B step.

Now, after a B step,
\begin{eqnarray}
1 - 2 p_Z' - 2 \Delta' & = & \left[1 - 2 p_Z^2 - 2 p_X (1 - 2 p_Z) - 
2 \Delta (1 - 2 p_X) \right] / p_S \\
& = & \left[2 p_Z (1 - p_Z) + (1 - 2 p_Z - 2 \Delta) (1 - 2 p_X)
\right] / p_S.
\end{eqnarray}
The first term is always positive, so the sum is clearly positive as
well when $1 - 2 p_Z - 2 \Delta > 0$.

After a P step,
\begin{equation}
1 - 2 p_Z' = (1 - 2 p_Z)^3,
\end{equation}
so
\begin{eqnarray}
1 - 2 p_Z' - 2 \Delta' & = & (1 - 2 p_Z)^3 - 6 \Delta^2 (1 - 2 p_Z) +
4 \Delta^3 \\ 
& = & (1 - 2 p_Z - 2 \Delta) \left[(1 - 2 p_Z)^2 + 
2 \Delta (1 - 2 p_Z - \Delta) \right].
\end{eqnarray}
Again, this is positive when $1 - 2 p_Z - 2 \Delta > 0$ and $\Delta \geq 
0$.  This proves the claim and the theorem.

\section{Inamori's Security Proofs}
\label{a:inamori}

In this Appendix, we provide a simple derivation of Inamori's proofs
of BB84 and the six-state scheme and discuss why our protocols
can tolerate a higher rate than his.

Inamori's protocols require two-way communications. His protocol
can be re-phrased as follows:

\begin{enumerate}

\item Alice and Bob are assumed to share initially a random
string and the goal of QKD is to extend this string.
Alice and Bob also choose a classical error correcting code $C_1$.

\item Alice sends Bob a sequence of single photons as in
either BB84 or the six-state scheme.

\item They throw away all polarization data that are prepared
in different bases and keep only the ones that are prepared in the
same bases.

\item They randomly select $m$ of those pairs and perform
a refined data analysis to find out the error rate of the various
bases. 

\item Alice measures the remaining $N-m=s $ particles
to generate a random string, $v$. Since $v$ is a random
string, it generally has non-trivial error syndrome when regarded as
a corrupted state of the codeword of $C_1$. Alice transmits that
error syndrome in an encrypted form to
Bob. This is done by using a one-time pad encryption with (part of)
the common string they initially share as the key.

\item Bob corrects his error to recover the string $v$.

\item Alice and Bob discard all the bits where they disagree and
keep only the ones where they agree.

\item Alice and Bob now perform privacy amplification on the
remaining string to generate a secure string.

\end{enumerate}

We remark that Inamori's protocol is, in fact, a simple
error correction scheme and satisfies the conditions of
Thm.~\ref{main}. Therefore, it is convenient to study it
using the language of two-way EPPs introduced in the current
extended abstract.

\subsection{BB84 with Inamori's protocol}
Let us now consider the efficiency of BB84 based on Inamori's
protocol. Suppose the error rate of each basis is found to be $p$
in step~4. Now,
in step~5 above, Alice and Bob have to sacrifice
a pre-shared secret key whose length
must be at least the size of the error syndrome of
an $s$-bit string. In other words, the length of the pre-shared
secret key used up by Alice and Bob is at least
\begin{equation}
l_{sac} = s h(p)
\label{e:lsac}
\end{equation}
bits where $h(x) = - x \log_2 x - ( 1 -x) \log_2 ( 1-x)$.

What is the length of the key they generate from the process?
Recall that in Step~7, Alice and Bob discard all the bits
where they disagree and keep only the ones where they agree.
The length of their reconciled key
is, therefore, given by the number of bits where Alice and
Bob agree. In other words, Alice and Bob generate a reconciled key of
the length
\begin{equation}
r= s ( 1- p) .
\label{e:recon}
\end{equation}

Since Eve may have some partial information on the reconciled key,
Alice and Bob have to sacrifice some of the reconciled key for
privacy amplification.
Let us consider privacy amplification.
For BB84, the worst case density matrix is again of the
form
\begin{equation}
diag ( 1 -2p, p, p,0)
\end{equation}
in the Bell-basis using the convention in \cite{BDSW}.

In step~7, Alice and Bob post-select only the bits where
they agree. With such post-selection, the
(unnormalized) conditional density matrix becomes:
\begin{equation}
diag ( 1 -2p, 0, p, 0).
\end{equation}
In other words, the phase error rate is:
\begin{equation}
{ p \over ( 1 - 2p+p)} = { p \over 1-p}.
\label{e:phasebb84}
\end{equation}

Therefore, Alice and Bob must sacrifice a further fraction
\begin{equation}
f_{BB84}= h \left( { p \over 1-p} \right)
\label{e:fractionbb84}
\end{equation}
of their reconciled key for privacy amplification.

In summary, the length of the reconciled key is $r= s ( 1-p)$, as given
by Eq.~(\ref{e:recon}). Of which, a fraction $h( { p \over 1-p})$
has to be consumed for privacy amplification. Therefore, the
final key generated by Alice and Bob is of length
$ [ 1- h( { p \over 1-p}) ] s ( 1-p) $.
In addition, from Eq.~(\ref{e:lsac}), a length of
$l_{sac} = s h(p)$ of a pre-shared secret key has to be
consumed. Therefore, the {\it net} key generation rate is
given by:
\begin{equation}
 \left[ 1- h \left( { p \over 1-p} \right) \right] s ( 1-p) -
s h(p) = s( 1-p)
 \left[ 1- h \left( { p \over 1-p} \right) - { h(p) \over 1-p } \right].
\label{e:nkgr}
\end{equation}

From Eq.~(\ref{e:nkgr}), one can conclude that in Inamori's protocol,
the net key generation rate is positive provided that:
\begin{equation}
 1 - h \left( { p \over 1-p}\right) - { h(p) \over 1 - p } >0,
\label{e:bb84+}
\end{equation}
which is exactly what appears just before Section~5 of \cite{inamori}.

Note that, for BB84, the maximal tolerable error rate of
Inamori's scheme is actually worse than in Shor-Preskill.

\subsection{Six-state scheme with Inamori's protocol}

Let us now consider the six-state scheme.
Suppose that in Step~4, the error rate is
found to be $p$. In Step~5, the length of
the pre-shared key sacrificed by Alice and Bob is the
same as in BB84 and is given by Eq.~(\ref{e:lsac}).
Also, the length of the reconciled key is the same as
in BB84 and is given by Eq.~(\ref{e:recon}).

Here is the key difference between the six-state scheme
and BB84: For the six-state scheme, there is more symmetry.
In particular, as discussed in Subsection~\ref{s:EPPproof},
for an EPP that corresponds to
the six-state scheme, one only needs to consider a
depolarizing channel.
The density matrix is:
\begin{equation}
diag ( 1- 3(p/2), p/2, p/2,p/2).
\end{equation}

On post-selecting the bits where Alice and Bob agree, the
(un-normalized) density matrix becomes:
\begin{equation}
diag ( 1 -3(p/2), 0, p/2, 0).
\end{equation}
Therefore, the post-selected phase error rate is:
\begin{equation}
 { p/2 \over 1 - 3(p/2) + p/2} = { p \over 2 ( 1- p) }.
\label{e:phasesix}
\end{equation}

Comparing Eqs.~(\ref{e:phasebb84}) and (\ref{e:phasesix}), we
see that a big difference between BB84 and six-state
in the Inamori's protocol is that the post-selected phase
error rate for the six-state is only half of that for BB84.
Consequently, Alice and Bob sacrifice fewer bits for privacy amplification
in the six-state case. In fact, only a smaller fraction, namely a fraction
\begin{equation}
f_{six} =
h \left(  { p \over 2 ( 1- p) } \right) 
\label{e:fractionsix}
\end{equation}
of the reconciled key needs to be sacrificed in the privacy amplification
process.

In summary, the length of the reconciled key is $r= s ( 1-p)$, as given
by Eq.~(\ref{e:recon}). Of which, from
Eq.~(\ref{e:fractionsix}), only a fraction $h( { p \over 2 (1-p) })$
has to be consumed for privacy amplification. Therefore, the
final key generated by Alice and Bob is of length
$ [ 1- h( { p \over 2 ( 1-p) } ) ] s ( 1-p) $.
In addition, from Eq.~(\ref{e:lsac}), a length of
$l_{sac} = s h(p)$ of a pre-shared secret key has to be
consumed. Therefore, the {\it net} key generation rate is
given by:
\begin{equation}
\left[ 1- h \left( { p \over 2( 1-p) } \right) \right] s ( 1-p)
- s h(p) = s( 1-p)
\left[ 1- h \left( { p \over 2( 1-p) } \right)- { h(p) \over 1-p}
\right].
\label{e:nkgrsix}
\end{equation}

From Eq.~(\ref{e:nkgrsix}), one can conclude that in Inamori's protocol
for the six-state scheme,
the net key generation rate is positive provided that:
\begin{equation}
 1 - h \left( { p \over 2( 1-p)}\right) - { h(p) \over 1 - p } >0,
\label{e:six+}
\end{equation}
which is precisely what Inamori gave in the Equation just under Property~1
on p. 3 of \cite{inamori6}.
Comparing Eqs.~(\ref{e:bb84+}) and (\ref{e:six+}),
one can see that the key difference between BB84 and six-state for
Inamori's protocol is in the second term of the expressions.
In the case of the six-state
scheme, there is an extra factor of $2$ in the
denominator inside the entropy function.
As noted before, this is because the six-state scheme has more
symmetry and gives a lower phase error rate (upon post-selection
of bits where Alice and Bob do agree) than BB84.

From Eq.~(\ref{e:six+}),
Inamori's protocol for the six-state
case can tolerate a bit error rate of roughly $12.6\%$.
A more recent protocol \cite{six} for the
six-state scheme can tolerate a marginally higher bit error rate and,
unlike Inamori's scheme, it requires only one-way classical
post-processing.  We remark that the six-state scheme with our
Protocol~2 tolerates a much higher error rate (about $23\%$, or as
high as $26.4\%$ varying the sequence of B and P steps) than a six-state
scheme with Inamori's protocol.

\end{document}